\documentclass[a4paper,12pt]{article}
\usepackage{etex}
\usepackage{epsf}
\usepackage[dvips]{graphicx}
\usepackage{a4wide}
\usepackage{amsmath}
\usepackage{amscd}
\usepackage{amsfonts}
\usepackage{natbib}
\usepackage{multirow}

\usepackage[sc]{mathpazo} 
\usepackage[T1]{fontenc} 

\usepackage{microtype} 
\usepackage[hang, small,labelfont=bf,up,textfont=it,up,skip=2.5pt]{caption}
\usepackage{booktabs}
\usepackage{lettrine}

\usepackage{abstract} 

\usepackage{titlesec} 
\titleformat{\section}[block]{\large\scshape\centering}{\thesection.}{1em}{} 
\titleformat{\subsection}[block]{\large}{\thesubsection.}{1em}{} 

\usepackage{titling} 
\usepackage{amssymb}
\usepackage{caption,subfig}
\usepackage{longtable}
\usepackage{rotating}
\usepackage{wrapfig}
\usepackage{pdflscape}
\usepackage{color}
\usepackage{multirow}
\usepackage{rotating}
\usepackage{natbib}
\usepackage{enumerate}
\usepackage{enumitem}
\usepackage{dsfont}
\usepackage{setspace}
\usepackage{anysize}
\usepackage{booktabs}
\usepackage{soul}
\usepackage{float}
\usepackage{enumitem}
\usepackage{tikz}
\usepackage{pifont}
\usepackage[hidelinks]{hyperref}
\usepackage{moresize}
\usetikzlibrary{arrows}
\usepackage{multirow}
\usepackage{ntheorem}
\usepackage{geometry}
\usepackage{tabularx}
\hypersetup{colorlinks=true,urlcolor=blue,linkcolor=blue,citecolor=blue}
\usepackage[section]{placeins}
\usepackage{textcomp}

\theoremstyle{}
\newtheorem{theorem}{Theorem}

\newtheorem{proposition}[theorem]{Proposition}
\newtheorem{corollary}{Corollary}[theorem]

\numberwithin{equation}{section}

\newenvironment{definition}[1][Definition]{\begin{trivlist}
\item[\hskip \labelsep {\bfseries #1}]}{\end{trivlist}}

\newcommand{\qed}{\nobreak \ifvmode \relax \else
      \ifdim\lastskip<1.5em \hskip-\lastskip
      \hskip1.5em plus0em minus1/2em \fi \nobreak
      \vrule height0.75em width1/2em depth0.25em\fi}
      
\marginsize{1.75cm}{1.75cm}{1.25cm}{1.25cm}
\long\def\symbolfootnote[#1]#2{\begingroup
\def\thefootnote{\fnsymbol{footnote}}\footnote[#1]{#2}\endgroup}

\DeclareGraphicsExtensions{.eps}

\date{}

\title{Identification of Peer Effects using Panel Data}
\author{Marisa Miraldo, Carol Propper, Christiern Rose\thanks{Miraldo and Propper: Imperial College Business School, Rose: University of Queensland. Corresponding author: Christiern Rose, School of Economics, University of Queensland, 4072, Australia, christiern.rose@uq.edu.au. This project is part of the Health Foundation's Efficiency Research programme. The Health Foundation is an independent charity working to improve the quality of healthcare in the UK. We are thankful for their financial support. Propper acknowledges financial support from the ERC. Hospital Episode Statistics datasets were provided by the Health and Social Care Information Centre (HSCIC, now National Health Service Digital). The Hospital Episode Statistics are copyright \textcopyright 2000/01-2013/14, re-used with the permission of The Health \& Social Care Information Centre. All rights reserved.}}

\renewcommand{\maketitlehookd}{%
\begin{abstract}
{\singlespacing \noindent We provide new identification results for panel data models with peer effects operating through unobserved individual heterogeneity. The results apply for general network structures governing peer interactions and allow for correlated effects. Identification hinges on a conditional mean restriction requiring exogenous mobility of individuals between groups over time. 
We apply our method to surgeon-hospital-year data to study take-up of keyhole surgery for cancer, finding a positive effect of the average individual heterogeneity of other surgeons practicing in the same hospital.

\noindent\textbf{Key words:} Peer effects, panel data, identification, networks, innovation, healthcare
}
\end{abstract}
}

\begin{document}

\maketitle


\section{Introduction}

This paper provides new identification results for panel data models of peer effects, through which outcomes depend on peers' unobservable (to the researcher) heterogeneity. Our framework also allows for \textit{correlated effects}, modelled as unobserved group heterogeneity which is permitted to be correlated with individual heterogeneity in an unrestricted manner.  We extend existing identification results to apply both to a general network structure (e.g., a social network) and to allow for correlated effects, and apply our results to study innovation take-up among physicians.

Identification depends on a conditional mean restriction requiring exogenous mobility of individuals between groups over time, as first formalised in \cite{abowd99}. That is, though our model of correlated effects allows, for example, that high outcome individuals be systematically located in high outcome groups, their mobility between groups ought not be determined by transitory outcome shocks. Not all patterns of mobility suffice for identification, and we provide identifying and non-identifying examples. We also provide an extension of our identification results to allow for endogenous peer effects, through which outcomes depend directly on peers' outcomes. With endogenous effects, identification can also be attained using the conditional mean restriction, however, for certain network structures, additional conditional variance restrictions are necessary. 
We conduct a Monte-Carlo with many individuals and few time periods and demonstrate that the NLS estimator first proposed by \cite{arci12} works well in practice. Increasing the number of time periods, the rate of mobility and the richness of the network (e.g., social network data) improves the performance of the estimator.  

Our empirical work considers innovation take-up in cancer treatment. The innovation we consider is keyhole surgery for colorectal cancer, and our data are from the English National Health Service. Colorectal cancer is the third most common cancer worldwide \citep{arnold17}. In England, it accounts for 10\% of cancer deaths and is the most expensive cancer to treat \citep{laudicella16}. Keyhole surgery for colorectal cancer is an important innovation. It has been shown to lead to better patient outcomes than the alternative open procedure, particularly in the short term. Moreover, in the National Health Service, keyhole surgery is less costly, primarily due to shorter post-surgery hospital stays by patients \citep{lacy02,nelson04,laudicella16}. Despite this, its take-up was slow, increasing from 1\% of eligible surgeries when it was first introduced in 2000 to 49\% by 2014. 

We use matched patient-surgeon-hospital-year data from 2000 to 2014 to estimate peer effects in take-up of keyhole surgery, measured by the fraction of colorectal cancer surgeries performed by keyhole. We find positive and statistically significant peer effects. Our results suggest that a standard deviation increase in the average latent take-up of other surgeons in the same hospital leads to a 5 percentage point increase in take-up. We decompose this effect by additionally estimating the effect of peer experience, which we find accounts for some, but not all, of the peer effect.

\subsection{Related Literature}

Research has largely focussed on settings in which peer effects operate through observable characteristics. That is, in addition to correlated effects, outcomes depend on  peers' observables \citep{manski93,moffitt01, lee07,bramoulle09,calvo09,davezies09,degiorgi10,goldsmith13,blume15,depaula17,cohen18,bramoulle19}. These papers consider a sample comprising a cross-section of groups, for which identification requires within-group variation in peers. Our panel data allow us to relax these requirements. First, the researcher need not observe individual characteristics, and if observed, need not know which ones are exogenous nor which ones are appropriate to include. Second, variation induced by mobility of individuals between groups over time means that there is no need for within-group variation in peers. This implies that identification is possible under the linear-in-means network structure, in which peer effects operate through the group average, and which precludes identification when only a cross-section of groups is available \citep{manski93,bramoulle09}. 

Another strand of the literature considers peer effects operating through individual unobservables using a cross-section of groups \citep{graham08,rose17}. Relative to these papers, we allow for unobserved group level heterogeneity to be arbitrarily correlated with individual heterogeneity, and, though our results can be applied to any network, we show that panel data can be used to identify peer effects for the linear-in-means network, which are not identifiable with only a cross-section of networks when there are correlated effects \citep{rose17}.

The most closely related work considers panel data models of peer effects. Key contributions are \cite{mas09}, \cite{arci12} and \cite{cornel17}, which, like us, study models of peer effects operating through the unobserved heterogeneity of peers. We build on their work by providing identification conditions which are straightforward to verify in practice, and which can be applied to a general network structure (e.g., social networks) in the presence of correlated effects. We also extend our results to the canonical model of endogenous peer effects, in which outcomes are simultaneously determined.\footnote{\cite{arci12} consider a variant of endogenous effects through which outcomes depend on the expected (as opposed to realized) outcomes of others, and study identification in an example with 2 individuals. This model does not allow simulataneity in outcomes.}


Beyond the peer effects literature our work can be viewed as extending the worker-firm fixed effects framework for wage decomposition used in labor economics (e.g., \cite{abowd99}) to allow for within-firm interactions of workers. That is, in addition to worker and firm heterogeneity, wages may depend on the composition of other workers in the firm as well as their wages. Such spillovers would be expected to operate in firms in which workers work in teams \citep{mas09,cornel17}.

Our empirical work constributes to the literature on innovation take-up, particularly in the healthcare context. Related papers are \cite{agha18} and \cite{barrenho20}. \cite{agha18} study how the take-up of new cancer drugs depends on the presence of local opinion leaders. To do this, they compare diffusion patterns across regions, separating correlated regional demand for new technology from information spillovers. They find that take-up is fastest in the region in which the lead author on the clinical trial practices. However, their work does not directly study peer effects in innovation. \cite{barrenho20} study the take-up of keyhole surgery, and estimate peer effects through surgeon observables. Our empirical contribution is to additionally allow peer effects to operate through surgeons' latent propensity to innovate, which we show matters above and beyond the effect of peer experience.

We proceed as follows. In Section \ref{model}, we present our baseline model. In Section \ref{ident} we provide identification results, apply them to two examples and discuss estimation. In Sections \ref{monte} and \ref{app} we conduct a Monte Carlo experiment and apply our method to surgeons' take-up of keyhole surgery for colorectal cancer. In Section \ref{con} we conclude. Proofs and extensions are in the Appendix.

\subsection{Notation}
If $M$ is a strictly positive integer, we denote $[M]=\{1,2,...,M\}$. If $\mathbf{A}$ and $\mathbf{B}$ are $M\times P$ and $M\times Q$ matrices, we denote the $M\times (P+Q)$ matrix obtained by concatenating $\mathbf{A}$ and $\mathbf{B}$ by $[\mathbf{A},\mathbf{B}]$. If element $(i,j)$ of $\mathbf{A}$ is $\mathbf{A}_{ij}$, we write $\mathbf{A}=(\mathbf{A}_{ij})_{i\in [M],j\in[P]}$, and if $\mathbf{a}$ is a vector, $\mathbf{a}_i$ denotes entry $i$. We use $\mathbf{I}_M$ for the $M$ dimensional identity, $\iota_M$ for the $M\times 1$ vector of ones and $\mathbf{0}$ to denote a matrix of zeros. If its dimensions are ambiguous we write $\mathbf{0}_{M,P}$ to denote a $M\times P$ matrix of zeros. We use $\mathbf{1}(\cdot)$ to denote the indicator function.
\section{Model}\label{model}

We consider a pattern of mobility of $N$ workers between $M$ groups over $T$ periods. Our interest lies in the typical setting in which $T$ is small and $N$ can be large. A mobility pattern is characterised by the $N\times M$ matrices of group membership indicators $\mathbf{C}_1,\mathbf{C}_2,...,\mathbf{C}_T$ and the $N\times N$ interaction matrices $\mathbf{G}_1,\mathbf{G}_2,...,\mathbf{G}_T$, the elements of which encode the peer effect exerted by one individual on another. The groups are such that in each period, every individual is in exactly one group and every group has at least one individual in at least one period. We use $g(i,t)\in[M]$ to denote the group of individual $i\in[N]$ in year $t\in[T]$ and $N_{g(i,t)}$ to denote its size. 

The $N\times 1$ vector of continuous outcomes in period $t\in[T]$ is $\bold{y}_t$, which is determined by 
\begin{align}
	\mathbf{y}_{t}&=\left(\mathbf{I}_N+\rho\mathbf{G}_t\right)\alpha+\mathbf{C}_t\gamma+\epsilon_{t}\label{stckt}
\end{align}
where $\alpha$ is an $N\times 1$ vector of time-invariant individual unobserved heterogeneity, $\gamma$ an $M\times 1$ vector of unobserved group heterogeneity, and $\epsilon_{t}$ an $N\times 1$ disturbance. In Section \ref{ext}, we consider an extension in which $\gamma$ is permitted to vary by period. The parameter of interest is $\rho$, which captures the peer effect.

Unless otherwise stated $\mathbf{G}_t$ is unrestricted and can be interpreted as the adjacency matrix of a weighted, directed network linking $N$ individuals. In particular, denoting entry $(i,j)$ of $\mathbf{G}_t$ by $\mathbf{G}_{ijt}$, it need not be the case that $\mathbf{G}_{ijt}=0$ when $g(i,t)\neq g(j,t)$, nor when $i=j$, nor when $\mathbf{G}_{jit}=0$. From this point onwards, we refer to $\mathbf{G}_t$ as the \textit{network}. A typical example of $\mathbf{G}_t$ is the \textit{linear-in-means} network, \begin{align}\overline{\mathbf{G}}_t=\left(\mathbf{1}(g(i,t)=g(j,t))N_{g(i,t)}^{-1}\right)_{(i,j)\in[N]^2}.\end{align} This implies that the peer effect operates through the group average of $\alpha_i$, and is a natural choice when only group membership indicators are available. If more detailed data on the network structure are available (e.g., social network data), this information can be incorporated into $\mathbf{G}_t$. Clearly, $\mathbf{G}_t$ can evolve over time as individuals move between groups. 

In our empirical application, $\mathbf{y}_t$ is surgeons' take-up of keyhole surgery for colorectal cancer, measured by the fraction of eligible surgeries performed by keyhole in year $t$, $\mathbf{C}_t$ comprises indicators for the hospital in which a surgeon practices, and we take $\mathbf{G}_t$ to be linear-in-means, linear-in-others'-means (in which the focal surgeon is excluded from the group average, see \eqref{liom}) and a persistent version of linear-in-others'-means, in which links persist when a surgeon moves to a new hospital and are weighted by the number of years worked at the same hospital. The latter captures cumulative peer exposure, which may be better suited to innovation take-up. The vector $\alpha$ captures surgeons' latent propensity to take up keyhole surgery. It can account for heterogeneity in education/training, ability (e.g., dexterity) and taste for innovation. The vector $\gamma$ captures hospital level heterogeneity including resources (e.g., equipment) and patient composition.

Stacking \eqref{stckt} by period yields $\mathbf{y}=\left(\mathbf{J}+\rho\mathbf{G}\right)\alpha+\mathbf{C}\gamma+\epsilon$, where $\mathbf{y}$ and $\epsilon$ are $NT\times 1$, $\mathbf{C}=(\mathbf{C}_1',\mathbf{C}_2',\hdots, \mathbf{C}_{T}')'$, $\mathbf{J}=(\mathbf{I}_N,\mathbf{I}_N,\hdots, \mathbf{I}_{N})'$ and $\mathbf{G}=(\mathbf{G}_1',\mathbf{G}_2',\hdots, \mathbf{G}_{T}')'$. Since $\sum_{i=1^N}\mathbf{J}_{ki}=\sum_{f=1}^M\mathbf{C}_{kf}=1$ for all $k\in[NT]$, we use the normalization $\gamma_M=0$ to obtain
\begin{align}
	\mathbf{y}=\left(\mathbf{J}+\rho\mathbf{G}\right)\alpha+\mathbf{D}\gamma+\epsilon\label{stcktt}
\end{align}
where $\mathbf{D}$ comprises the first $M-1$ columns of $\mathbf{C}$ and from this point forwards $\gamma=(\gamma_1,\gamma_2,...,\gamma_{M-1})'$. This is without consequence for identification of $\rho$.

Our identification results depend on variation in the network (both over individuals and over time) and mobility of individuals between groups over time.
It is well known that mobility serves to separate the individual and correlated effects (e.g., \cite{abowd99}). As we show below, mobility also serves to separate individual and correlated effects from the peer effect. This is because, in the typical setting in which there are no between group links,\footnote{i.e., if $g(i,t)\neq g(j,t)$ then $\mathbf{G}_{ijt}=0$, though our results do not require this.} if an individual moves from one group to another she ceases to interact with others in her previous group and begins to interact with others in her new group. 

\section{Identification}\label{ident}

Our identification results treat the joint distribution of $\mathbf{y},\mathbf{G},\mathbf{D}$ as observable. This is consistent with the researcher accessing a sample from this distribution. For small $T$, our approach is identical in spirit to that used throughout the peer effects literature (i.e., with $T=1$), in which the researcher is assumed to observe a cross-section of groups (see \cite{bramoulle19} for a review). 

A more challenging but sometimes more realistic alternative is that the researcher observes a sample of individuals from a single group, so that, depending on the network structure, all individuals are potentially linked to one another, at least indirectly. It is more challenging because additional structure is required to deal with the dependence between individuals, though this is primarily a concern for inference rather than identification. \cite{goldsmith13} describe how asymptotic analysis could be implemented based on a random variable measuring `distance' between individuals. Their argument is as follows. If distant individuals have low probability of link formation (e.g., due to homophily), and distance has large support, it may be possible to construct blocks of individuals such that each pair of blocks could be treated as close to independent. This is in the same spirit as certain types of asymptotic analysis for time-series, and allows the researcher to view a sample of individuals from a single group similarly to a sample of many groups. 

\cite{goldsmith13} use the above arguments to justify conducting identification analysis as if the joint distribuion of $\mathbf{y},\mathbf{G},\mathbf{X}$ were observable, where $\mathbf{X}$ is a matrix of exogenous characteristics through which peer effects may operate.\footnote{Their model does not include correlated effects, hence the absence of $\mathbf{D}$.} Their arguments can be directly applied in our context to justify analysis of a sample from the joint distribution of $\mathbf{y},\mathbf{G},\mathbf{D}$ if $T$ is small. The only caveat is that $\mathbf{y},\mathbf{G},\mathbf{D}$ must include all observed time periods for the individuals and groups, so that temporal dependence is not an issue in (hypothetically) constructing blocks of observations. In our application, such blocks could be thought of as corresponding to regions of England because mobility is primarily between hospitals in the same region \citep{goldacre13,barrenho20}. In other applications such as wage decomposition in labor economics, the relevant partition could be by industry and/or by region. 

We study identification of $\rho$ based on the conditional mean restriction
\begin{align}
\mathbb{E}[\mathbf{y}|\mathbf{G},\mathbf{D}]=\left(\mathbf{J}+\rho\mathbf{G}\right)\mathbb{E}[\alpha|\mathbf{G},\mathbf{D}]+\mathbf{D}\mathbb{E}[\gamma|\mathbf{G},\mathbf{D}],\label{mom}
\end{align}
which implies exogeneity of the network and mobility of individuals between groups over time with respect to the outcome shock $\epsilon$. The former is typical in the peer effects literature and the latter is typical in the wage decomposition literature. We do not restrict $\mathbb{E}[\alpha|\mathbf{G},\mathbf{D}]$ nor $\mathbb{E}[\gamma|\mathbf{G},\mathbf{D}]$, which allows for a limited form of network endogeneity with respect to unobserved individual and group heterogeneity.

We say that $\rho$ is identified when it can be uniquely recovered from the right-hand side of \eqref{mom}. Our results are thus asymptotic in nature (see \cite{manski95}), and hence charaterize whether peer effects can be distentangled from individual and group heterogeneity if there is no limit to the number of mobility patterns observed. To simplify the exposition, following \cite{bramoulle09} and \cite{abowd99}, the remainder of the paper presents the case in which $\mathbf{G}$ and $\mathbf{D}$ are treated as fixed. To allow for the random case, we simply replace unconditional expectations with expectations conditional on $\mathbf{G},\mathbf{D}$, and the identification results below hold if there exists a realization in the support of $\mathbf{G},\mathbf{D}$ which satisfies the relevant condition. In the same way, it is straightforward to allow $N,M$ and $T$ to vary by mobility pattern. Identical arguments are used throughout the peer effects literature (see, e.g., \cite{bramoulle09}). Returning to the fixed case, \eqref{mom} becomes
\begin{align}
\mathbb{E}[\mathbf{y}]=\left(\mathbf{J}+\rho\mathbf{G}\right)\mu^\alpha+\mathbf{D}\mu^\gamma,\label{mom1}
\end{align}
where $\mu^\alpha=\mathbb{E}[\alpha]$ and $\mu^\gamma=\mathbb{E}[\gamma]$.  Since \eqref{mom1} is non-linear in the parameter $\rho$ and the (unknown) $\mu^\alpha$, establishing identification is non-trivial, depending both on the properties of the $NT\times 2N+M-1$ matrix $[\mathbf{J},\mathbf{G},\mathbf{D}]$ and on the value of $\mu^\alpha$. 
For example, it is clear that $\rho$ is not identified when $\mu^\alpha=\mathbf{0}$. 

Mobility of individuals between groups over time is necessary for identification. In the absence of mobility, $[\mathbf{J},\mathbf{G},\mathbf{D}]$ has rank at most $N$, so \eqref{mom1} yields $N$ equations in $N+M$ unknowns. For the same reason, we also require $T\geq 2$. This is because the peer effect operates through time-invariant individual unobserved heterogeneity, rather than through individual observables. However, mobility alone is not sufficient for identification, as made clear in Example 2 below.

We now present our first identification result, which makes use of the within-group annihilator for the correlated effects, given by $\mathbf{W}=\mathbf{I}_{NT}-\mathbf{D}(\mathbf{D}'\mathbf{D})^{-1}\mathbf{D}$, and a decomposition of vectors $\mathbf{v}=(\mathbf{v}_1',\mathbf{v}_2')'$ which lie in the null-space of $[\mathbf{WJ},\mathbf{WG}]$ such that $\mathbf{v}_1$ and $\mathbf{v}_2$ are both $N\times 1$.
\begin{proposition}\label{the1}
$\rho$ is identified if there does not exist a vector $\mathbf{v}=(\mathbf{v}_1',\mathbf{v}_2')'$ in the null-space of $[\mathbf{WJ},\mathbf{WG}]$ and scalars $\delta_1$ and $\delta_2\neq 0$ verifying $\delta_1\mathbf{v}_1+\delta_2\mathbf{v}_2=\mu^\alpha$. Otherwise $\rho$ is not identified. 
\end{proposition}
\noindent Notice that full column rank of $[\mathbf{J},\mathbf{G},\mathbf{D}]$ is not necessary because $\mathbf{v}=\mathbf{0}$ need not be the only vector in the null-space of $[\mathbf{WJ},\mathbf{WG}]$. This is because $[\mathbf{J},\mathbf{G},\mathbf{D}]$ has $2N+M-1$ columns but there are only $N+M$ unknowns. Requiring $[\mathbf{J},\mathbf{G},\mathbf{D}]$ to have full column rank is too strong because it rules out $T=2$.\footnote{This is because full column rank requires $NT\geq 2N+M-1$, hence $T\geq2+(M-1)/N$. $T=2$ is not immediately ruled out when $M=1$, but $\rho$ is not identifiable in this case because there can be no mobility if there is only 1 group.}
A common empirical setting is when the rows of $\mathbf{G}$ sum to one (e.g., linear-in-means), such that peer effects operate through a weighted average. If there are no other collinearities among the columns of $[\mathbf{J},\mathbf{G},\mathbf{D}]$ then one can apply the following.
\begin{corollary}\label{co2}
 If $\mathbf{G}\iota_{N}=\iota_{NT}$, $\rho$ is identified if ${\rm rank}([\mathbf{J},\mathbf{G},\mathbf{D}])=2N+M-2$ and there exists $(i,j)\in[N]^2$ such that $\mu^\alpha_i\neq\mu^\alpha_j$.
\end{corollary}
Corollary \ref{co2} can be shown using the decomposition $[\mathbf{WJ},\mathbf{WG}]=\mathbf{S}\mathbf{R}$ where $\mathbf{S}$ is the $NT\times 2N-1$ full rank matrix formed by concatenating $\mathbf{WJ}$ and the first $N-1$ columns of $\mathbf{WG}$ and
\begin{align}
\mathbf{R}=(\mathbf{S}'\mathbf{S})^{-1}\mathbf{S}'[\mathbf{WJ},\mathbf{WG}]=\begin{pmatrix}
\mathbf{I}_N&\mathbf{0}_{N,N-1}&\iota_N\\
\mathbf{0}_{N-1,N}&\mathbf{I}_{N-1}&-\iota_{N-1}
\end{pmatrix}.
\end{align}
From the structure of $\mathbf{R}$, it is immediate that vectors $\mathbf{v}$ in the null-space of $[\mathbf{WJ},\mathbf{WG}]$ are of the form $\mathbf{v}_1=c\iota_N$, $\mathbf{v}_2=-\mathbf{v}_1$ for $c\in\mathbb{R}$. Applying Proposition \ref{the1} yields identification of $\rho$ if there exists $(i,j)\in[N]^2$ such that $\mu^\alpha_i\neq\mu^\alpha_j$. If this condition is violated then $\mu^\alpha=a\iota_N$ for some $a\in\mathbb{R}$, and $(\mathbf{J}+\rho\mathbf{G})\mu^\alpha=(1+\rho)a\iota_{NT},$ so only $(1+\rho)\mu^\alpha$ is identifiable. Intuitively, we require  $\mu^\alpha_i\neq\mu^\alpha_j$ because if individuals are homogeneous then no amount of mobility can lead to changes in the average of $\alpha_i$ over peers, hence outcomes do not vary in response to changes in peer composition over time.


The identification conditions above depend on $\mu^\alpha$, which is not observed. We now ask how `large' is the set of values of $\mu^\alpha$ for which $\rho$ is identified. This is the notion of generic identification (see \cite{lewbel19}). If $\rho$ is generically identified, then it is identified for all values of $\mu^\alpha$ with the exception of a few pathological cases, which are `unlikely' to arise in practice. 
\begin{corollary}\label{co0}
If ${\rm rank}([\mathbf{WJ},\mathbf{WG}])\geq N+1$ the set of $\mu^\alpha$ for which $\rho$ is not identified is a measure zero subset of $\mathbb{R}^N$.
\end{corollary}
Corollary \ref{co0} means that $\rho$ is generically identified if ${\rm rank}([\mathbf{WJ},\mathbf{WG}])\geq N+1$. This is because $\mathbf{v}$ lies in a subspace of $\mathbb{R}^{2N}$ of dimension at most $N-1$, hence $\delta_1\mathbf{v}_1+\delta_2\mathbf{v}_2$ lies in a subspace of $\mathbb{R}^N$ of dimension at most $N-1$. 


As with the well known rank condition in linear models (e.g., no perfect multicollinearity among regressors for linear regression), the researcher can check whether the rank requirements of Corollaries \ref{co2} and \ref{co0} hold in the observed data. To do this, one interprets $N$ as the total number of observed individuals, $M$ as the total number of observed firms and $T$ as the total number of observed time periods, and constructs the observed values of $\mathbf{J},\mathbf{G}$ and $\mathbf{D}$ accordingly.\footnote{In our identification analysis, these quantities are the number of individuals, firms and time periods in a single mobility pattern (i.e., a draw from the joint distribution of $\mathbf{y},\mathbf{G},\mathbf{D}$). The observed data comprise the realizations of many such patterns.} If the rank requirement of either Corollary holds then the researcher can be confident of identification. 

Proposition \ref{the1} can be viewed as panel data analogue of the identification results of \cite{bramoulle09}, which imply that, for $T=1$ and exogenous observed characteristic(s) $\mathbf{X}$ of dimension $N\times K$, the peer effect $\dot\rho$ in the model \begin{align}\mathbb{E}[\mathbf{y}|\mathbf{X}]=\mathbf{X}\dot\beta+\mathbf{GX}\dot\rho+\mathbf{D}\dot\mu^\gamma\label{brr}\end{align} is identified if and only if $\mathbf{WJ}$ and $\mathbf{WG}$ are linearly independent (i.e., if there does not exist nonzero $\lambda\in\mathbb{R}^2$ such that $\lambda_1\mathbf{WJ}+\lambda_2\mathbf{WG}=\mathbf{0}$).\footnote{Since $\alpha$ does not appear in \eqref{brr}, there is no need to impose $\gamma_M=0$, which is no longer a normalization. In this case $\mathbf{D}$ ought to be replaced by $\mathbf{C}$ and $\mathbf{W}$ by the annihilator for $\mathbf{C}$ when referring to the model in \eqref{brr}. We continue to use the notation $\mathbf{D}$ and $\mathbf{W}$ to facilitate the comparison with our results.} This is a weaker rank requirement than that in Proposition \ref{the1}, and can be satisfied when $T=1$. This is because the peer effect is assumed to operate through the observed $\mathbf{X}$ rather than through the unobserved $\alpha$. In practice the researcher may not know what to include in $\mathbf{X}$ and/or $\mathbf{X}$ may be of large dimension and/or endogenous. Proposition \ref{the1} shows that identification can be attained with $T\geq 2$ without requiring any knowledge on $\mathbf{X}$.\footnote{Of course this requires that $\mathbf{X}$ be time-invariant. However, common choices of the components of $\mathbf{X}$ such as gender and education are also time-invariant.} An additional advantage of panel data is that it facilitates identification for some network structures for which peer effects are not otherwise identifiable. For example, if $T=1$ and the network is linear-in-means, then $\mathbf{WG}=\mathbf{0}$ and \eqref{brr} does not identify the peer effect. In contrast, as we show in Example 1 below, peer effects are identifiable when $T=2$ provided that there is mobility of individuals between groups over time. 

It is straightforward to extend our model to include exogenous characteristics, yielding
\begin{align}
\mathbb{E}[\mathbf{y}|\mathbf{X}]=\mathbf{X}\beta+\mathbf{F}\mathbf{X}\rho_1+\left(\mathbf{J}+\rho_2\mathbf{G}\right)\mu^\alpha(\mathbf{X})+\mathbf{D}\mu^\gamma(\mathbf{X})\label{rhoX}
\end{align}
where $\mathbf{X}$ is $NT\times K$, $\mathbf{F}$ is a $NT\times NT$ block diagonal matrix with blocks $\mathbf{G}_1,\mathbf{G}_2,...,\mathbf{G}_T$, $\mu^\alpha(\mathbf{X})=\mathbb{E}[\alpha|\mathbf{X}]$ and $\mu^\gamma(\mathbf{X})=\mathbb{E}[\gamma|\mathbf{X}]$. Provided that the entries of $\mathbf{X}$ vary over time, the parameters $\rho_1$ and $\rho_2$ are separately identifiable under similar conditions to Proposition \ref{the1}. For brevity, we do not pursue this formally, though we do estimate such a specification in our empirical application. 

In the Appendix we provide analagous idenification results which additionally allow for endogenous peer effects, through which outcomes are simultaneously determined. We also provide conditional variance restrictions in the spirit of \cite{graham08} and \cite{rose17}, which can be used when the conditional mean restriction does not suffice for identification. 

We now consider two examples of our baseline identification results with $N=M=T=2$.\\ 

\noindent \textbf{Example 1: An identifying mobility pattern.} Consider the following mobility pattern in which individuals one and two are respectively in groups 1 and 2 in the first period. In the second period, individual one remains in group 1 and individual two moves from group 2 to group 1. Under the linear-in-means network, this mobility pattern yields
\begin{align}
[\mathbf{J},\mathbf{G},\mathbf{D}]=
\begin{pmatrix}
    1&      0&       1 &      0&1\\
    0&      1&       0 &      1&0\\
    1&      0&       1/2&    1/2&1\\
    0&      1&       1/2&    1/2&1\\
\end{pmatrix},\quad
[\mathbf{WJ},\mathbf{W}\mathbf{G}]=\begin{pmatrix}1/3&-1/3&1/3&-1/3\\0 &1&0&1\\1/3&-1/3&-1/6&1/6\\-2/3&2/3&-1/6&1/6 \end{pmatrix},
\end{align}
which respectively have rank $4$ and $3$.  Since $[\mathbf{J},\mathbf{G},\mathbf{D}]$ has rank $2N+M-2$, by Corollary \ref{co2}, 
$\rho$ is identified if $\mu^\alpha_1\neq\mu^\alpha_2$. Since ${\rm rank}[\mathbf{WJ},\mathbf{W}\mathbf{G}]=N+1$, Corollary \ref{co0} states that $\rho$ is generically identified. This is because the subset of $\mathbb{R}^2$ such that $\mu^\alpha_1=\mu^\alpha_2$ has measure zero.

For the intuition, consider the underlying system of equations for the outcomes $y_{it}$ of individual $i$ in period $t$,
\begin{align}
\begin{array}{ll}
\mathbb{E}[y_{11}]=\mu^\alpha_1+\rho\mu^\alpha_1+\mu^\gamma_1 &\mathbb{E}[y_{21}]=\mu^\alpha_2+\rho\mu^\alpha_2\\
\mathbb{E}[y_{12}]=\mu^\alpha_1+\rho(\mu^\alpha_1+\mu^\alpha_2)/2+\mu^\gamma_1&\mathbb{E}[y_{22}]=\mu^\alpha_2+\rho(\mu^\alpha_1+\mu^\alpha_2)/2+\mu^\gamma_1\label{xc}
\end{array}
\end{align}
When individual two moves groups, individual one obtains a new peer, hence the peer effect on individual one changes from $\rho\mu^\alpha_1$ in the first period to $\rho(\mu^\alpha_1+\mu^\alpha_2)/2$ in the second period, whilst the individual and correlated effect are unchanged. The change due to the peer effect is given by the expected change in the outcome of individual one between the first and second period $\mathbb{E}[y_{11}]-\mathbb{E}[y_{12}]=\rho(\mu^\alpha_1-\mu^\alpha_2)/2$. To identify $\rho$ we now need to identify $\mu^\alpha_1-\mu^\alpha_2$. We can use the second period, in which both individuals are in the same group, hence have the same correlated and peer effects. This means that any difference in their expected outcomes is due to differences in their individual effects, so $\mathbb{E}[y_{12}]-\mathbb{E}[y_{22}]=\mu^\alpha_1-\mu^\alpha_2$. If $\mu^\alpha_1\neq\mu^\alpha_2$ we obtain \begin{align}\rho=2(\mathbb{E}[y_{11}]-\mathbb{E}[y_{12}])/(\mathbb{E}[y_{12}]-\mathbb{E}[y_{22}]).\end{align} 

If we were to observe this mobility pattern repeatedly, we could estimate the expectations using sample means, yielding an estimator of $\rho$. Of course, in practice we do not repeatedly observe the same mobility pattern, but a variety of patterns, the information from which we combine through an estimator based on the conditional mean restriction \eqref{mom}. Nevertheless, for the purposes of identification we require only that there exists a single identifying mobility pattern realized with non-zero probability. Finally, notice that the above arguments are unchanged when the normalization $\gamma_M=0$ is not used, in which case one has $\mathbb{E}[y_{21}]=\mu^\alpha_2+\rho\mu^\alpha_2+\mu^\gamma_2$ in \eqref{xc}.\\

\noindent \textbf{Example 2: A non-identifying mobility pattern.} Now modify Example 1 such that both individuals move in the second period. This means that the individuals are in different groups in the first period and swap groups in the second period, implying that $\mathbf{G}=\mathbf{J}$. The null-space of $[\mathbf{WJ},\mathbf{W}\mathbf{G}]$ comprises $\mathbf{v}=(\mathbf{v}_1',\mathbf{v}_2')'=(\mathbf{u}',-\mathbf{u}')'$ for any $\mathbf{u}\in\mathbb{R}^2$. For any value of $\mu^\alpha\in\mathbb{R}^2$, there clearly exists $\mathbf{u}\in\mathbb{R}^2$ and scalars $\delta_1$ and $\delta_2\neq 0$ such that $(\delta_1-\delta_2)\mathbf{u}=\mu^\alpha$, so by Proposition \ref{the1}, $\rho$ is not identified. Note also that the identification condition in Corollary \ref{co2} is violated since  $[\mathbf{J},\mathbf{G},\mathbf{D}]$ has rank $3<2N+M-2=4$ and the generic identification condition in Corollary \ref{co0} is violated since $[\mathbf{WJ},\mathbf{W}\mathbf{G}]$ has rank $2<N+1=3$. 

For the intuition, consider again the underlying system of equations
\begin{align}
\begin{array}{ll}
\mathbb{E}[y_{11}]=\mu^\alpha_1+\rho\mu^\alpha_1+\mu^\gamma_1 &\mathbb{E}[y_{21}]=\mu^\alpha_2+\rho\mu^\alpha_2\\
\mathbb{E}[y_{12}]=\mu^\alpha_1+\rho\mu^\alpha_1&\mathbb{E}[y_{22}]=\mu^\alpha_2+\rho\mu^\alpha_2+\mu^\gamma_1
\end{array}
\end{align}
The correlated effect $\mu^\gamma_1$ is identified by individual one moving groups ($\mu^\gamma_1=\mathbb{E}[y_{11}]-\mathbb{E}[y_{12}]$) but the remaining equations are only sufficient to identify $(1+\rho)\mu^\alpha$. The reason for this is that there is no variation in the peers of either individual because the individuals are never in the same group in the same period. Mobility of individuals between groups over time is insufficient for identification of $\rho$ because it does not induce changes in peer groups. This contrasts with the canonical wage decomposition model imposing $\rho=0$, for which group-swapping would be sufficient to identify $\mu^\alpha_1,\mu^\alpha_2,\mu^\gamma_1$.\footnote{Note that identification of $\mu^\alpha$ and $\mu^\gamma$ is only up to the normalization $\gamma_M=0$.} Since it does not identify $\rho$, no matter how many times we observe this mobility pattern in our data, it cannot be used to construct an estimator of $\rho$.

\subsection{Time-varying Correlated Effects}\label{ext}

If correlated effects are time-varying, then $\gamma_{g(i,t)}$ is replaced by $\gamma_{g(i,t)t}$, in which case $\mathbf{C}$ is a $NT\times MT$ block diagonal matrix with blocks $\mathbf{C}_1,\mathbf{C}_2,...,\mathbf{C}_T$, $\mathbf{D}$ comprises the first $MT-1$ columns of $\mathbf{C}$ and $\gamma$ is $(MT-1)\times 1$. All Propositions then apply as stated, provided that $\mathbf{W}$ is modified accordingly. The parameters are not identifiable under the linear-in-means network because the peer effect varies only at the group-period level, so cannot be separated from the correlated effect. 

\subsection{Estimation}\label{est}

\cite{arci12} propose NLS estimation of \eqref{stcktt}, treating $\alpha,\rho,\gamma$ as parameters to be estimated (i.e., a fixed effects approach). The authors study its properties under the \textit{linear-in-others'-means} network,
\begin{align}
\widetilde{\mathbf{G}}_t=\left(\mathbf{1}(g(i,t)=g(j,t),i\neq j)(N_{g(i,t)}-1)^{-1}\right)_{(i,j)\in[N]^2},\label{liom}
\end{align} 
and without correlated effects. Assuming that $\mathbb{E}[\epsilon_{it}\epsilon_{js}]=0$ for all $i\neq j, t\neq s$, $\mathbb{E}[\epsilon_{it}^2|g(i,t)]=\mathbb{E}[\epsilon_{jt}^2|g(j,t)]$ for all $i,j,t$ such that $g(i,t)=g(j,t)$, $\mathbb{E}[\epsilon_{it}\alpha_j]=0$ for all $i,j,t$ and $\rho<\min_{i,t}N_{g(i,t)}$, the authors show that the estimator of $\rho$ is consistent and asymptotically normal in the number of individuals provided that there are at least two time periods. The authors discuss a variant of correlated effects appropriate to their application to peer effects in education, which is allowed to vary over time but is restricted to be the same across multiple groups, and argue that they expect similar behavior of the estimator in this case. 
The proposed NLS estimator is based on the conditional mean restriction \eqref{mom}, hence, subject to identification, could equally be applied to other network structures and specifications of the correlated effect. We do not formally establish its properties because our focus is on identification and our empirical application. However we do explore this in our Monte-Carlo experiment.

\section{Monte Carlo Experiment}\label{monte}

The design is tailored to our empirical application, with 700 individuals, 140 groups, 15 time periods, mobility rate $p=0.03$ (this is the probability that an individual moves groups from one period to the next, see summary statistics in Table \ref{sumstat}) and $\rho=0.5$. We also consider designs with 2 time periods and mobility rate $p=0.1$. 

The data generating process is as follows. In the first period, all individuals are randomly assigned to groups of size five. In each subsequent period, each individual moves group with probability $p$, in which case she draws a new group with uniform probability over all groups. The expected group size is 5 for all groups in all periods. The network structures we consider are linear-in-means, linear-in-others'-means, and a variant of linear-in-others'-means in which links persist when individuals move groups and are weighted by the number of periods spent in the same group, and a social network, in which there is within-group variation in peers. 

For the persistent network, we let $\mathbf{A}_s$ be the $N\times N$ binary adjacency matrix in period $s$, with element $(i,j)$ equal to 1 if $g(i,s)=g(j,s)$ and $i\neq j$. Then we define $\mathbf{G}_t$ by taking $\sum_{s=1}^t\mathbf{A}_s$ and rescaling its rows to sum to 1. This means that $\mathbf{G}\alpha$ captures both contemporaneous and cumulative exposure to others. 

The social network is constructed as follows. In the first period, each individual draws two links uniformly over other individuals in the same group. In each subsequent period, links persist whilst individuals remain in the same group. If an individual loses a link(s) due to mobility, a replacement link(s) is drawn uniformly over the other individuals in the group with whom there is not already a link. Links need not be reciprocal.\footnote{If an individual is in a group of size 1 they have no link.} If there exists a link between $i$ and $j$ in period $t$ then $\mathbf{G}_{ijt}$ is equal to the inverse of the number of links that $i$ has in period $t$. Otherwise  $\mathbf{G}_{ijt}=0$. If there are three or fewer individuals in the group, each individual is linked to all other individuals. 

We take $\alpha_i=1+W_i+\eta_i$, where $W_i\in[0,1]$ is the number of moves made by individual $i$ divided by $(T-1)$ and $\eta_i\sim\mathcal{N}(0,1)$. We set $\epsilon_{it}\sim\mathcal{N}(0,1/2)$ and consider cases in which $\gamma=\mathbf{0}$ (which is also imposed on the estimator) and in which $\gamma_m$ is the mean of $\alpha_i$ over all members of group $m$ and all time periods. This design means that high $\alpha_i$ individuals are more mobile and tend to be located in high $\gamma_m$ groups. We choose the variances of $\alpha_i$ and $\epsilon_{it}$ as above so that, on average, a standard deviation increase in the peer effect on individual $i$ in period $t$ (i.e., in $\sum_{j=1}^N\mathbf{G}_{ijt}\alpha_j$) leads to a 0.15-0.3 (depending on the network structure) standard deviation increase in $y_{it}$. This matches the effect sizes we find in our empirical application. We simulate 500 datasets for each experiment. Every dataset verifies the generic identification condition in Corollary \ref{co0}.

\begin{table}[t!]
\centering \caption{Monte Carlo Results}\label{resultsmc0}
{\small
\begin{tabular}{l | c c c c c c c c}
\hline\hline
&\multicolumn{8}{c}{Unobserved $\alpha$}\\
\hline
LIM	&0.501&0.5&0.523&0.522&0.502&0.503&0.496&0.515\\
	&(0.04)&(0.051)&(0.297)&(0.456)&(0.026)&(0.032)&(0.161)&(0.447)\\
LIOM	&0.5&0.5&0.505&0.454&0.501&0.502&0.494&0.437\\
	&(0.029)&(0.035)&(0.216)&(0.355)&(0.019)&(0.022)&(0.121)&(0.386)\\
PLIOM& 0.501&0.507&0.505&0.394&0.503&0.503&0.489&0.455\\
	&(0.068)&(0.085)&(0.414)&(0.641)&(0.046)&(0.052)&(0.231)&(0.361)\\
SOC&0.501&0.5&0.5&0.462&0.5&0.5&0.503&0.496\\
	&(0.019)&(0.023)&(0.185)&(0.361)&(0.013)&(0.014)&(0.084)&(0.248)\\
\hline
&\multicolumn{8}{c}{Observed $\alpha$}\\
\hline
LIM	& 0.5&0.5&0.498&0.493&0.501&0.502&0.499&0.491\\
	&(0.009)&(0.037)&(0.027)&(0.175)&(0.009)&(0.024)&(0.026)&(0.12)\\
LIOM	&0.5&0.499&0.499&0.495&0.5&0.502&0.499&0.493\\
	&(0.008)&(0.028)&(0.022)&(0.139)&(0.008)&(0.018)&(0.021)&(0.096)\\
PLIOM&0.5&0.501&0.499&0.497&0.501&0.501&0.499&0.49\\
	&(0.008)&(0.03)&(0.022)&(0.165)&(0.008)&(0.024)&(0.021)&(0.115)\\
SOC&0.5&0.499&0.5&0.498&0.5&0.5&0.5&0.498\\
	&(0.007)&(0.013)&(0.019)&(0.039)&(0.007)&(0.012)&(0.017)&(0.039)\\

\hline
\hline
Correlated Effects&No&Yes&No&Yes&No&Yes&No&Yes\\
Periods&15&15&2&2&15&15&2&2\\
Mobility Rate ($p$)&0.03&0.03&0.03&0.03&0.1&0.1&0.1&0.1\\
Individuals&700&700&700&700&700&700&700&700\\
Groups&140&140&140&140&140&140&140&140\\
\hline\hline
\end{tabular}}\\
\footnotesize{\textbf{Notes:} We report the mean and standard deviation (in parentheses) estimate of $\rho$ over 500 datasets, with true value $0.5$.  'LIM' is the linear-in-means network, `LIOM' is linear-in-others'-means, `PLIOM' is the persistent LIOM and `SOC' is the social network. `Unobserved $\alpha$' is the NLS estimator in Section \ref{est}. `Observed $\alpha$' is the OLS estimator taking $\mathbf{X}=\mathbf{J}\alpha$ in \eqref{brr}. If there are no correlated effects the estimator imposes $\gamma=\mathbf{0}$. Correlated effects are time-invariant.}
\end{table}

\subsection{Results}

Table \ref{resultsmc0} reports the results. The top panel reports the NLS estimator of $\rho$ described in Section \ref{est}. To provide a benchmark for comparison, the bottom panel reports the infeasible OLS estimator that would be used if $\alpha$ were observable (i.e., for the model in \eqref{brr} taking $\mathbf{X}=\mathbf{J}\alpha$). Columns 1-2  of Table \ref{resultsmc0} show designs with 15 periods and mobility rate 0.03, which match our empirical application. The NLS estimator of $\rho$ is centered on the true value over all networks and with and without correlated effects. The variance of the NLS estimator is of a similar order of magnitude to the infeasible OLS estimator, though of course it is larger. Columns 3-4 reduce the number of periods to 2, which causes an increase in the variance of the estimator. The final 4 columns increase the rate of mobility to 0.1, which reduces the variance of the estimator. Comparing the rows of Table \ref{resultsmc0}, we find that peer effects are most precisely estimated for the social network. This is likely because the social network exhibits the most within-group variation in peers.

\section{Application}\label{app}

We study surgeons' take-up of keyhole surgery for colorectal cancer in the English National Health Service (henceforth, NHS). As discussed in the introduction, colorectal cancer is prevalent, costly to treat, and accounts for a high proportion of cancer deaths worldwide. An important innovation in its treatment is keyhole surgery, which reduces costs and improves patient outcomes relative to the older open procedure. Despite this, take-up of keyhole surgery in England was slow, increasing from 1\% of eligible surgeries in 2000, the year it was introduced in England, to 49\% by 2014 (see Table \ref{sumstat}). Our goal is to study the extent to which its diffusion was driven by peer effects, hence by mobility of surgeons between hospitals over time.

The NHS is an ideal setting for our empirical work because it treats almost all cancer patients in England,\footnote{There is a small private sector in England that mainly provides care for planned procedures for which there are long waiting lists. Private sector provision for (any) cancers during the period we examine was very limited and primarily focused on treatment of overseas patients.} it is a public system in which surgeons are salaried employees of one hospital at any point in time (hence their renumeration does not depend on the treatment provided), all hospitals operate under the same financial rules set by central government and have the technology necessary for keyhole surgery, and a two week waiting time guarantee for cancer referral implies that allocation of patients to surgeons is close to random, based on surgeon availability in a local hospital \citep{barrenho20}. 

\subsection{Data}

Our data are from \cite{barrenho20}, comprising a panel of NHS surgeons and their take-up of keyhole surgery from its introduction in 2000 through to 2014. The data are obtained by merging Hospital Episode Statistics, which provides treatment information for all patients in the English NHS with NHS Workforce Statistics and the General Medical Council register. This provides matched patient-surgeon-hospital-year data, which is collapsed into a surgeon-hospital-year panel. 
To be included in the estimation sample, surgeons must be observed at least twice, must perform more than 5 colorectal cancer surgeries,\footnote{This is to avoid including surgeons who do not routinely perform the procedure.} 
and hospital-year pairs must have at least two surgeons. The resulting unbalanced panel comprises 11,923 observations of 1,363 surgeons over 15 years across 194 hospitals. 

The dependent variable is surgeon take-up of keyhole surgery for colorectal cancer, measured by the fraction of eligible colorectal cancer sugeries performed by keyhole in the focal year.\footnote{Keyhole surgery is suitable for some, but not all, patients. The sample of patients considered is restricted to those for which the surgeon has a choice between keyhole and the alternative open surgery. A detailed description of patient eligibility can be found in \cite{barrenho20}.} The surgeons we observe are senior physicians, known as Consultants. They are entirely autonomous, and, for the patients we consider, have discretion to choose either keyhole or open surgery. We also observe surgeon demographics (gender and age, though we omit gender from our models because it is time-invariant), and compute an experience measure which is the cumulative number of eligible colorectal cancer surgeries performed (both keyhole and open) by the beginning of the focal year. To account for increasing average take-up over time, we also include year fixed effects in all specifications.

We apply our model for surgeon $i$ in hospital $g(i,t)$ in year $t$,\footnote{We set a surgeon's hospital to be that at which they practiced for the majority of days of the year.} and consider the linear-in-means and linear-in-others'-means networks, as well as the persistent linear-in-others'-means network used in our Monte-Carlo experiment. Recall that in the persistent network, a surgeon is a peer if they have ever worked concurrently in the same hospital, and peers are weighted by number of years worked together, with weights summing to one. This network captures both contemporaneous and cumulative exposure to others, hence measures a `stock' of peer influence, rather than the `flow' measured by the other networks.

Table \ref{sumstat} summarises the data. In a typical hospital in a typical year there are 5 surgeons performing colorectal cancer surgery. The largest has 16 surgeons. A typical surgeon is in their early 40s and has performed around 140 colorectal cancer surgeries by the beginning of a typical year. We only observe surgeons beginning in 2000, hence experience is equal to zero in this year for all surgeons. The minimum value of experience can be zero in any year due to entry of newly qualified surgeons.\footnote{For this reason, in addition to taste and ability, $\alpha$ may be thought of as including experience in colorectal cancer surgery prior to 2000.}

\begin{table}[t!]
\centering \caption{Summary Statistics}\label{sumstat}
{\small
\begin{tabular}{l l | c c c c c c c c c}
\hline\hline
Year&&2000&2003&2006&2009&2012&2014&All\\
\hline\hline
\multicolumn{2}{l|}{Surgeons observed}&600   &      761    &     782   &      834      &   891    &     854      &  1363\\
\hline
\multirow{4}{*}{Take-up}&Mean&0.008   &    0.022 &      0.104    &   0.293   &    0.421   &    0.485  &     0.218\\
								&S.D.&0.031     &  0.072    &   0.173   &    0.273   &     0.28  &    0.291  &      0.28 \\
								&Min& 0   &  0   &  0   &  0    & 0   &  0     &0  \\
								&Max&0.256   & 1   & 1   & 1   & 1  &  1   & 1 \\
								\hline
Age $<40$&Mean&0.179    &    0.19   &    0.201   &    0.225  &     0.207    &   0.252   &    0.233\\
Age $40-44$&Mean&0.283      & 0.246   &    0.253  &     0.233    &   0.284   &    0.284    &    0.26\\
Age $45-49$&Mean& 0.227     &  0.172   &    0.239  &     0.194  &     0.219    &   0.219  &     0.212\\
Age $50-54$&Mean&0.196&  0.153 &      0.165 &      0.131 &      0.166   &     0.16    &   0.165\\
\hline
\multirow{4}{*}{Experience (00's)}&Mean&0    &   0.689    &   1.279    &   1.771      & 2.104     &  2.486   &    1.418\\
								&S.D.& 0     &  0.548    &   1.028     &  1.444     &  1.838     &  2.013    &    1.51\\
								&Min& 0      &     0     &      0     &      0     &      0    &    0.01     &      0 \\
								&Max&0      &  2.65     &   5.01     &   6.65      &  8.62    &    9.37    &    9.37\\
\hline								
Moved						    &Mean& -   &    0.035    &   0.028   &    0.038     &  0.035   &    0.023   &    0.029\\
Peers changed				&Mean&-    &   0.582    &   0.602    &    0.64    &   0.738    &   0.542   &    0.601\\
\hline										   
\hline
\multicolumn{2}{l|}{Hospitals observed}&151 & 151  &149 & 148 & 143&  139 & 194\\
\hline
\multirow{4}{*}{No. of surgeons}    &Mean&3.974     &   5.04   &    5.248  &     5.635     &  6.231   &    6.144      &  5.36\\
										    &S.D. &1.689    &   2.343   &    2.339   &    2.419    &   2.533      & 2.541   &    2.439\\
										    &Min  & 2  &   2  &   2  &   2  &   2   &  2  &   2\\
										    &Max &10 & 14 & 12 & 13 & 16  &14 & 16\\
\hline\hline
\end{tabular}}\\
\footnotesize{\textbf{Notes:} `Take-up' is the fraction of colorectal cancer surgeries performed by keyhole. `Age' are indicators for the specified range. `Experience` is 00's of colorectal cancer surgery at the beginning of the year. `Moved' is a binary indicator for a surgeon being located in a different hospital in year $t$ than in year $t-1$. `Peers changed' is a binary indicator for a surgeons' peers being different in year $t$ than in $t-1$. `No. of surgeons' is the number of surgeons located in a hospital.}
\end{table}


\subsection{Identification}

Our identification results are based on a sample from the joint distribution of $\mathbf{y},\mathbf{G},\mathbf{D}$. As argued in Section \ref{ident}, we also expect them to be applicable if the sample can be partitioned into blocks based on individuals' `distance' to one another, such that the dependence between each pair of blocks is limited. Mobility of surgeons between hospitals is largely within the local region \citep{goldacre13,barrenho20}, hence such blocks could (hypothetically) be constructed by partitioning regions. Due to this, we expect that our theory provides a reasonable approximation.

Identification is based on exogenous mobility of surgeons between hospitals over time. We expect this to be the case because all hospitals have the required technology for keyhole surgery, the NHS is a public system with similar working conditions and renumeration nationwide, and colorectal cancer forms only a small part of surgeons' workloads. The leading reason for mobility is to relocate closer to the pre-medical school family home \citep{goldacre13}. \cite{barrenho20} provide empirical evidence of exogenous mobility, showing that past take-up does not predict mobility, and that, conditional on moving, there is no association between take-up of the surgeon and take-up of the hospital moved to. These arguments support the conditional mean restriction in \eqref{mom}. Since $\mathbb{E}[\alpha|\mathbf{G},\mathbf{D}]$ and $\mathbb{E}[\gamma|\mathbf{G},\mathbf{D}]$ are unrestricted, we allow for dependence between a surgeon's latent propensity to take-up keyhole surgery and the nature of their network. Moreover, we do not rule out high take-up surgeons being systematically located in high take-up hospitals, though we do rule out their mobility between hospitals being driven by transitory take-up shocks. 

We now discuss the extent of identifying variation in the data. Though only 3\% of surgeons move hospital in a typical year (see Table \ref{sumstat}), our panel is relatively long, comprising 15 years in total. Moreover, each move changes the peer groups of all those in the hospital left and all those in the hospital joined. The median number of surgeons in a hospital-year pair is 5, hence a surgeon moving from one median sized hospital to another changes the peer groups of 10 surgeons. For this reason, 60\% of surgeons' peers change in a typical year. These changes in peer groups can generate large fluctuations in average peer take-up because peer groups are small. Our Monte-Carlo experiment also demonstrates that the observed mobility rate is sufficient to accurately estimate $\rho$. 

All estimated models verify generic identification of $\rho$. In the baseline model without surgeon demographics, the rank of $[\mathbf{WJ},\mathbf{WG}]$ is 2361 for the linear-in-means network, 2678 for the linear-in-others'-means network and 2683 for the persistent network. Since there are $N=1363$ surgeons, the rank requirement of Corollary \ref{co0} is satisfied.\footnote{For these calculations, year dummies are included to construct the annihilator $\mathbf{W}$, hence the partialling out is with respect to both hospital and year dummies. For models with surgeon demographics, these are also included to construct $\mathbf{W}$.}

\subsection{Results}

\begin{table}[t!]
\centering \caption{Peer Effects in Keyhole Surgery for Colorectal cancer}\label{results}
{\small
\begin{tabular}{l | c c c| c c c| c c c}
\hline\hline
&\multicolumn{3}{c|}{Linear-in-means}&\multicolumn{3}{c|}{Linear-in-others'-means}&\multicolumn{3}{c}{Persistent LIOM}\\
\hline
$\mathbf{G}\alpha$												&0.427&0.531&0.422&0.564&0.58&0.507&0.993&0.995&0.99\\
																	&(0.06)&(0.061)&(0.072)&(0.036)&(0.036)&(0.045)&(0.008)&(0.008)&(0.013)\\
																	\hline
Age$<40$															&&&-0.004&&&0.001&&&0.022\\
																	&&&(0.022)&&&(0.023)&&&(0.021)\\
Age$40-44$														&&&0.03&&&0.035&&&0.047\\
																	&&&(0.017)&&&(0.018)&&&(0.017)\\
Age$45-49$														&&&0.044&&&0.042&&&0.049\\
																	&&&(0.012)&&&(0.013)&&&(0.012)\\
Age$50-54$														&&&0.03&&&0.029&&&0.032\\
																	&&&(0.008)&&&(0.008)&&&(0.008)\\
Exper.																&&&0.043&&&0.047&&&0.045\\
																	&&&(0.003)&&&(0.003)&&&(0.003)\\
	\hline																
$\mathbf{F}$Age $<40$										&&&-0.003&&&-0.012&&&0.019\\
																	&&&(0.023)&&&(0.019)&&&(0.027)\\
$\mathbf{F}$Age $40-44$										&&&0.01&&&0.007&&&0.022\\
																	&&&(0.019)&&&(0.016)&&&(0.02)\\
$\mathbf{F}$Age $45-49$										&&&-0.014&&&-0.008&&&0.005\\
																	&&&(0.02)&&&(0.017)&&&(0.017)\\
$\mathbf{F}$Age $50-54$										&&&0.002&&&0.003&&&0.013\\
																	&&&(0.017)&&&(0.013)&&&(0.011)\\
$\mathbf{F}$Exper.												&&&0.025&&&0.035&&&0.008\\
																	&&&(0.008)&&&(0.006)&&&(0.005)\\

\hline	
$\rho\left(\sum_{i=1}^N\mathbf{G}_{ijt}^2\right)^{1/2}$&0.18&0.122&0.103&0.269&0.193&0.161&0.443&0.345&0.354\\
$\sigma_\alpha$												&0.168&0.253&0.294&0.181&0.262&0.315&0.207&0.297&0.306\\
\hline
Surgeon FE	 &Yes&Yes&Yes&Yes&Yes&Yes&Yes&Yes&Yes\\
Hospital FE	&No&Yes&Yes&No&Yes&Yes&No&Yes&Yes\\
Year FE	 &Yes&Yes&Yes&Yes&Yes&Yes&Yes&Yes&Yes\\
\hline
Observations&11923&11887&11887&11923&11887&11887&11923&11887&11887\\
\hline\hline
\end{tabular}}\\
\footnotesize{\textbf{Notes:} To allow for heteroskedasticity, standard errors are computed using a wild bootstrap, as proposed by \cite{arci12}. $\sigma^2_\alpha$ is computed using the estimated $\alpha$, as proposed by \cite{arci12}.}
\end{table}

Table \ref{results} reports the results. The first row reports estimates of $\rho$. We find positive peer effects in all specifications. The peer effect is precisely estimated, and statistically distinguishable from zero in all specifications. This is in line with our Monte-Carlo results, and, though their application to peer effects in education is different to ours, with the results of \cite{arci12}.

The vector $\alpha$ combines both surgeons' ability and taste for innovation, but we cannot decompose the two with the available data. Clearly, the size of the peer effect is interpretable only relative to the variance of $\alpha$. To provide an interpretable benchmark, we treat $\mathbf{G}$ as fixed and $\alpha_i$ as i.i.d., with variance $\sigma^2_\alpha$ estimated using the estimated $\alpha$, as proposed by \cite{arci12}. For surgeon $i$ in year $t$,  the effect on take-up of a standard deviation increase in average peer latent propensity to innovate is $\rho\sigma_\alpha\left(\sum_{i=1}^N\mathbf{G}_{ijt}^2\right)^{1/2}$. For the linear-in-means network this is $\rho\sigma_\alpha N_{g(i,t)}^{-1/2}$, and for linear-in-others'-means we replace $N_{g(i,t)}$ by $N_{g(i,t)}-1$. Since the effect of a standard deviation increase in own latent propensity to innovate is $\sigma_\alpha$, the magnitude of the peer effect relative to the own effect is $\rho\left(\sum_{i=1}^N\mathbf{G}_{ijt}^2\right)^{1/2}$. Table \ref{results} reports its average over all observations. With hospital fixed effects, it is around 0.1 for linear-in-means, 0.15-0.2 for linear-in-others'-means and 0.35 for the persistent network. Our estimates imply that a standard deviation increase in average peer latent propensity to innovate increases take-up by 3 percentage points for linear-in-means, 5 percentage points for linear-in-others'-means and 10 percentage points for the persistent network. Our finding that the persistent network corresponds to the largest effect size suggests that both contemporaneous and cumulative peer exposure play a role.

For each network, we also estimate a specifications which includes the age and experience of a surgeon and their peers (i.e., the model in \eqref{rhoX}). We find an inverted-U profile in age, with take-up estimated to peak between 45 and 49. The coefficient on experience is positive, suggesting that an additional hundred career colorectal cancer surgeries leads to around a 5 percentage point increase in take-up. The effect of peer age is close to zero, whilst peer experience plays a similar role to own experience. The estimate of $\rho$ is smaller than in models in which peer effects operate through $\alpha$ alone, though it remains positive and statistically significant at conventional levels. This suggests that some, but not all, of the peer effect operates through experience.

\section{Conclusion}\label{con}

This paper provides and implements new identification results for panel data models with peer effects. Our results suggest that these can typically be separated from correlated effects provided either that there is sufficient mobility in the data or that the network data are sufficiently detailed. We find positive peer effects in surgeons' take-up of keyhole surgery for colorectal cancer, which operate through peer experience and latent propensity to take-up the innovation. Our results imply that exposure to others with high take-up and experience may be useful to increase innovation take-up. Based on this, policymakers might consider programmes which expose low take-up surgeons to high take-up and/or experienced surgeons. For example, one might conceive a targeted secondment programme.

{\singlespacing \small
\bibliographystyle{ier}
\bibliography{refs}
}
\section*{Appendix}

\renewcommand{\theequation}{A.\arabic{equation}}

\noindent\textbf{Proof of Proposition \ref{the1}}

\noindent Denote $\theta=(\mu^\alpha,\rho,\mu^\gamma)$. Equation \eqref{mom1} yields $\mathbb{E}[\mathbf{Wy}]=\mathbf{WJ}\mu^\alpha+\mathbf{WG}\rho\mu^\alpha$. 
Suppose that there is $\overline\theta$ satisfying the conditional moment restriction in \eqref{mom1}. Then we have
\begin{align}
[\mathbf{WJ},\mathbf{WG}]\begin{pmatrix}\mu^\alpha-\overline\mu^\alpha\\ \rho\mu^\alpha-\overline\rho\overline\mu^\alpha \end{pmatrix}=\mathbf{0}
\end{align} 
Let $\mathbf{v}=(\mathbf{v}_1',\mathbf{v}_2')$ be a vector in the null-space of $[\mathbf{WJ},\mathbf{WG}]$. Then for some $\mathbf{v}$ we have
\begin{align}
\mathbf{v}_1=\mu^\alpha-\overline\mu^\alpha,\quad \mathbf{v}_2=\rho\mu^\alpha-\overline\rho\overline\mu^\alpha
\end{align}
which implies $\overline\rho\mathbf{v}_1-\mathbf{v}_2=(\overline\rho-\rho)\mu^\alpha$. If $\overline\rho\neq\rho$ then this is equivalently expressed as
\begin{align}
\frac{\overline\rho}{\overline\rho-\rho}\mathbf{v}_1-\frac{1}{\overline\rho-\rho}\mathbf{v}_2=\mu^\alpha
\end{align}
If there does not exist $\mathbf{v}$ in the null-space of $[\mathbf{WJ},\mathbf{WG}]$ and scalars $\delta_1$ and $\delta_2\neq 0$ verifying $\delta_1\mathbf{v}_1+\delta_2\mathbf{v}_2=\mu^\alpha$ then, by contradiction we have $\overline\rho=\rho$, so $\rho$ is identified. Otherwise we can clearly have $\overline\rho\neq\rho$, so $\rho$ is not identified.

\subsection*{Endogenous peer effects}

We now consider an extension of our identification results to models with endogenous peer effects. We consider the canonical model, in which outcomes are simutaneously determined (see \cite{bramoulle19} for a review of endogenous peer effects). The outcome equation is
\begin{align}
	&y_{it}=\alpha_i+\psi\sum_{j=1}^N\mathbf{G}_{ijt}y_{jt}+\rho\sum_{j=1}^N\mathbf{G}_{ijt}\alpha_j+\gamma_{g(i,t)}+\epsilon_{it},
\end{align}
or in stacked form,
\begin{align}
	\mathbf{y}=\psi\mathbf{F}\mathbf{y}+(\mathbf{J}+\rho\mathbf{G})\alpha+\mathbf{D}\gamma+\epsilon,
\end{align}
where $\psi$ is a scalar parameter capturing the endogenous peer effect and $\mathbf{F}$ is a $NT\times NT$ block diagonal matrix with blocks $\mathbf{G}_1,\mathbf{G}_2,...,\mathbf{G}_T$. 

Throughout this section we suppose that the rows of $\mathbf{G}$ sum to one ($\mathbf{G}\iota_{N}=\iota_{NT}$), that there are no between-group interactions ($\mathbf{G}_{ijt}=0$ if $g(i,t)\neq g(j,t)$) and $|\psi|<1$. These are standard assumptions maintained in almost all papers concerning identification of peer effects, and are made to ensure that the reduced form exists and the reduced form correlated effect is proportional to the structural correlated effect. Solving for the reduced form and taking expectations yields
\begin{align}\mathbb{E}[\mathbf{y}]=(\mathbf{I}_{NT}-\psi\mathbf{F})^{-1}(\mathbf{J}+\rho\mathbf{G})\mu^\alpha+\mathbf{D}(1-\psi)^{-1}\mu^\gamma.\end{align} We now modify Proposition \ref{the1} to allow for endogenous effects, making use of $\mathbf{H}=((\mathbf{G}_1^2)',(\mathbf{G}_2^2)',...,(\mathbf{G}_T^2)')'$ and vectors $\mathbf{v}=(\mathbf{v}_1',\mathbf{v}_2',\mathbf{v}_3')'$ in the null-space of $[\mathbf{WJ},\mathbf{WG},\mathbf{WH}]$ such that $\mathbf{v}_1,\mathbf{v}_2,\mathbf{v}_3$ are all $N\times 1$.
\begin{proposition}\label{the1y}
$\psi$ and $\rho$ are identified if $\psi+\rho\neq 0$ and there does not exist a vector $\mathbf{v}=(\mathbf{v}_1',\mathbf{v}_2',\mathbf{v}_3')'$ in the null-space of $[\mathbf{WJ},\mathbf{WG},\mathbf{WH}]$ and scalars $\delta_1,\delta_2,\delta_3,\delta_4$ satisfying $\delta_1\mathbf{v}_1-\mathbf{v}_2=(\delta_1-\delta_2)\mu^\alpha$, $\delta_3\mathbf{v}_1+\mathbf{v}_3=(\delta_3-\delta_4)\mu^\alpha$ and either $\delta_1\neq \delta_2$ or $\delta_3\neq\delta_4$. Otherwise, $\psi$ and $\rho$ are not identified. 
\end{proposition}
The identification conditions are similar in form to those of Proposition \ref{the1}. Similarly to Corollary \ref{co2}, generic identification of $\rho$ and $\psi$ is attained whenever $\mathbf{v}$ lies in a subspace of $\mathbb{R}^N$ of dimension strictly less than $N$. A sufficient condition which can be used in practice is that $[\mathbf{WJ},\mathbf{WG},\mathbf{WH}]$ has rank at least equal to $2N+1$. The additional requirement that $\psi+\rho\neq 0$ is needed because $\psi+\rho=0$ implies $\mathbb{E}[\mathbf{y}]=\mathbf{J}\mu^\alpha+\mathbf{D}(1-\psi)^{-1}\mu^\gamma,$ such that $\psi$ and $\rho$ are not identifiable. This is because the endogenous effect exactly offsets the contextual effect, yielding no net peer effect. 

As with Proposition \ref{the1}, Proposition \ref{the1y} can also be related to the work of \citep{bramoulle09}. The authors results imply that for $T=1$, the peer effects $\dot\psi$ and $\dot\rho$ in the model
 \begin{align}\mathbb{E}[\mathbf{y}|\mathbf{X}]=(\mathbf{I}_{N}-\dot\psi\mathbf{F})^{-1}(\mathbf{X}\dot\beta+\mathbf{GX}\dot\rho)+\mathbf{D}(1-\dot\psi)^{-1}\dot\mu^\gamma\end{align} are identified if and only if $\mathbf{WJ},\mathbf{WG}$ and $\mathbf{WH}$ are linearly independent (i.e., if there does not exist nonzero $\lambda\in\mathbb{R}^3$ such that $\lambda_1\mathbf{WJ}+\lambda_2\mathbf{WG}+\lambda_3\mathbf{WG^2}=\mathbf{0}$) and $\dot\beta\dot\psi+\dot\rho\neq 0$. Analagously to the discussion of Proposition \ref{the1}, the linear independence requirement is weaker and can be satisfied with $T=1$ because $\mathbf{X}$ is assumed to be observable and exogenous. These requirements can be relaxed when $T\geq 2$. The condition $\dot\beta\dot\psi+\dot\rho\neq 0$ is analagous to $\psi+\rho\neq 0$ in Proposition \ref{the1y}, and also rules out the case of no net peer effect.

Under the linear-in-means network, $\rho$ and $\psi$ are not separately identifiable without additional restrictions. This is because $\mathbf{H}=\mathbf{G}$, and we have
\begin{align}
\mathbb{E}[\mathbf{y}]=\left(\mathbf{J}+\frac{\psi+\rho}{1-\psi}\overline{\mathbf{G}}\right)\mu^\alpha+\mathbf{D}(1-\psi)^{-1}\mu^\gamma\label{Elim}
\end{align}
so that only the reduced form parameter $(\psi+\rho)(1-\psi)^{-1}$ is identifiable. Since \eqref{Elim} takes the same form as \eqref{mom1}, the results of Section \ref{ident} can be applied to establish identification of $(\psi+\rho)(1-\psi)^{-1}$.

To separately identify $\rho$ and $\psi$, one can use additional conditional variance restrictions. To this end, we consider the variance 
\begin{align}
\mathbb{V}[\mathbf{Wy}]=&\mathbf{W}(\mathbf{I}_{NT}-\psi\mathbf{F})^{-1}(\mathbf{J}+\rho\mathbf{G})\mathbb{V}[\alpha](\mathbf{J}+\rho\mathbf{G})'(\mathbf{I}_{NT}-\psi\mathbf{F}')^{-1}\mathbf{W}\nonumber\\
+&\mathbf{W}(\mathbf{I}_{NT}-\psi\mathbf{F})^{-1}\mathbb{V}[\epsilon](\mathbf{I}_{NT}-\psi\mathbf{F}')^{-1}\mathbf{W}\nonumber
\end{align}
and the restrictions
\begin{align}
\mathbb{COV}[\epsilon_{it},\epsilon_{js}]=\begin{cases}\sigma^2(g(i,t))&i=j,t=s\\ 0&\text{otherwise}  \end{cases},\quad\mathbb{COV}[\alpha_i,\alpha_j]=\begin{cases}\sigma^2_\alpha&i=j\\ 0&\text{otherwise} \end{cases}.\label{var}
\end{align}
As with the conditional mean restriction in \eqref{mom} considered thus far, when $\mathbf{G}$ and $\mathbf{D}$ are treated as random the variance and restrictions above are conditional on $\mathbf{G},\mathbf{D}$. We continue to present the fixed case for ease of exposition.

Our approach is similar in spirit to \cite{graham08} and \cite{rose17}, which consider a cross-section of networks. The restriction on the variance of $\epsilon$ requires that the transitory shocks experienced by individuals in the same group be uncorrelated and have equal variance. Uncorrelatedness implies that outcomes are correlated only due to the correlated effect and peer effects, whereas equality of variance implies that within a group, no individuals are subject to systematically larger shocks than others. Importantly, $\mathbb{COV}[\alpha_i,\gamma_{g(j,t)}]$ is unrestricted for all $i,j$ and $t$. This allows for sorting of individuals to groups based on their time-invariant heterogeneity. 

We first present the identification result for the linear-in-means network structure. In the next section we derive an identification result which applies to general network structures and under weaker conditional variance restrictions. In general, we require only that $\mathbb{COV}[\epsilon_{it},\epsilon_{js}]=0$ if $g(i,t)\neq g(j,t)$. The stronger restriction in \eqref{var} is necessary for the linear-in-means network, which is well known to be the most challenging case \citep{manski93,bramoulle09}.
\begin{proposition}\label{the2y}
If $\mathbf{G}=\overline{\mathbf{G}}$, $\psi$ and $\rho$ are identified if $\mathbf{W},\mathbf{WJJ'W},\mathbf{W}(\mathbf{JG}'+\mathbf{GJ}')\mathbf{W},\mathbf{WGG'W}$ and $\mathbf{WFW}$ are linearly independent.
\end{proposition}
In contrast to Proposition \ref{the1y} there is no requirement that $\psi+\rho\neq 0$. This is because the endogenous effect operates through the outcomes, and hence propagates variation both in $\alpha$ and $\epsilon$, whilst the contextual effect operates through $\alpha$ alone. Since they operate through different channels, the peer effects can be separated under restrictions on the within-group variance of $\epsilon$ such as those in \eqref{var}. The identification condition in Proposition \ref{the2y} fails if $T=1$ because $\mathbf{F}=\mathbf{G}=\mathbf{G}'=\mathbf{GG}'$ and $\mathbf{J}=\mathbf{I}_N$ so that $\mathbf{W}(\mathbf{JG}'+\mathbf{GJ}')\mathbf{W}=2\mathbf{WGG'W}=2\mathbf{WFW}$. This is in agreement with \cite{rose17}, which shows that the conditional variance restrictions in \eqref{var} are insufficient for identification with a cross-section of networks.\footnote{\cite{graham08} shows that contextual effects can be identified with a cross-section of networks under additional restrictions, including on $\mathbb{COV}[\alpha_i,\gamma_{g(j,t)}]$.}

Due to its simplistic nature, the identification condition in Proposition \ref{the2y} fails in Example 1. However, it typically holds in more realistic examples. For example, identification is restored if a third individual is added and individual three is in group 2 in both periods.

\subsubsection*{Conditional Variance Restrictions for a General Network Structure}

In the general network case identification can be attained without making restrictions on the within-group variance of $\epsilon$. This is in contrast to the linear-in-means network.\footnote{See the end of the proof of Proposition \ref{the2y} for an explanation of why restrictions on the within-group variance are necessary for the linear-in-means model.} One can use 
\begin{align}
\mathbb{COV}[\epsilon_{it},\epsilon_{js}]=0\quad g(i,t)\neq g(j,s),\quad
\mathbb{COV}[\alpha_i,\alpha_j]=\begin{cases}\sigma^2_\alpha&i=j\\ 0&\text{otherwise} \end{cases} \label{vargen}
\end{align}
instead of \eqref{var}, which allows for unrestricted within-group variance of $\epsilon$. We make use of the following definition from \cite{rose17}, reprinted here for convenience.
\begin{definition} \textit{Consider $L$ matrices of the same dimension $\mathbf{A}_1,...,\mathbf{A}_L$. The matrix $\mathbf{A}_l$ ($l\in[L]$) is maximally linearly independent of $\mathbf{A}_1,...,\mathbf{A}_L$ if $\lambda_l=0$ for all $\lambda\in\mathbb{R}^L$ such that $\sum_{l=1}^L\lambda_l\mathbf{A}_l=\mathbf{0}$.}
\end{definition}
Note that linear independence of $\mathbf{A}_1,...,\mathbf{A}_L$ is equivalent to each of the $L$ matrices being maximally linearly independent. In what follows we require only that a strict subset be maximially linearly independent, which is weaker than linear independence. Identification hinges on the covariance terms for outcomes of observations in different groups, which are non-zero provided that there is mobility between groups. To extract these covariance terms, we use $\mathcal{E}_m\subseteq[NT]$ to denote the indices of the rows of $\mathbf{y}$ corresponding to group $m\in[M]$ and define $\mathbf{E}_m$ as the matrix constructed from rows $\mathcal{E}_m$ of $\mathbf{I}_{NT}$. Pre-multiplying any conformable matrix by $\mathbf{E}_m$ extracts rows $\mathcal{E}_m$, and post-multiplying by $\mathbf{E}_m'$ extracts columns $\mathcal{E}_m$. We also use $\mathbf{E}_{-m}$ to extract the rows in the complement of $\mathcal{E}_m$.
\begin{proposition}\label{the3y}
$\psi,\rho,\sigma^2_\alpha$ are identified if $\psi+\rho\neq 0$ and there exists $m\in[M]$ such that $\mathbf{E}_{m}\mathbf{WJJ'W}\mathbf{E}_{-m}',  \mathbf{E}_m\mathbf{W}(\mathbf{JG}'+\mathbf{GJ}')\mathbf{W}\mathbf{E}_{-m}'$ and $\mathbf{E}_m\mathbf{W}(\mathbf{JH}'+\mathbf{HJ}')\mathbf{W}\mathbf{E}_{-m}'$ are maximially linearly independent from $\mathbf{E}_m\mathbf{WJJ'W}\mathbf{E}_{-m}',\mathbf{E}_m\mathbf{W}(\mathbf{JG}'+\mathbf{GJ}')\mathbf{W}\mathbf{E}_{-m}',\mathbf{E}_m\mathbf{W}(\mathbf{JH}'+\mathbf{HJ}')\mathbf{W}\mathbf{E}_{-m}',\mathbf{E}_m\mathbf{W}(\mathbf{GH}'+\mathbf{HG}')\mathbf{W}\mathbf{E}_{-m},\mathbf{E}_m\mathbf{WGG'W}\mathbf{E}_{-m}',$ and $\mathbf{E}_m\mathbf{WHH'W}\mathbf{E}_{-m}'$.
\end{proposition}
Identically to Proposition \ref{the1y}, $\psi+\rho\neq 0$ is required so that the endogenous and contextual effects do not exactly offset one another. This contrasts with Proposition \ref{the2y}, which concerns the linear-in-means network, for which $\psi+\rho\neq 0$ is not required. The reason for this is that Proposition \ref{the2y} additionally restricts the within-group variance of $\epsilon$ because restrictions on the between-group variance alone are insufficient for identification. Since the endogenous effects operate both through $\epsilon$ and $\alpha$, whereas the contextual effects operate only through $\alpha$, under restrictions on the within-group variance of $\epsilon$ there are no values of $\psi$ and $\rho$ such that the two peer effects exactly offset one another in the reduced form variance.

\noindent\textbf{Proof of Proposition \ref{the1y}}

\noindent Denote $\theta=(\psi,\rho,\alpha,\gamma)$. Under the conditional moment restriction we have
\begin{align}
\mathbb{E}[\mathbf{y}_t]=(\mathbf{I}_N-\psi\mathbf{G}_t)^{-1}(\mathbf{I}_N\mu^\alpha+\mathbf{G}_t\rho\mu^\alpha+\mathbf{D}_t\mu^\gamma)
\end{align}
for $t\in[T]$. The inverse of $(\mathbf{I}_N-\psi\mathbf{G}_t)$ exists because $\mathbf{G}_t\iota_N=\iota_N$ and $|\psi|<1$. Since $\mathbf{G}_t\iota_N=\iota_N$ and $\mathbf{G}_{ijt}=0$ if $g(i,t)\neq g(j,t)$ we have $(\mathbf{I}_N-\psi\mathbf{G}_t)^{-1}\mathbf{D}_t\mu^\gamma=\mathbf{D}_t(1-\psi)^{-1}\mu^\gamma$, hence
\begin{align}
\mathbb{E}[\mathbf{y}_t]=(\mathbf{I}_N-\psi\mathbf{G}_t)^{-1}(\mathbf{I}_N+\rho\mathbf{G}_t)\mu^\alpha+\mathbf{D}_t(1-\psi)^{-1}\mu^\gamma
\end{align}
Suppose that there is $\overline\theta$ satisfying the conditional moment restriction. Then we have
\begin{align}
(\mathbf{I}_N-\psi\mathbf{G}_t)^{-1}(\mathbf{I}_N+\rho\mathbf{G}_t)\mu^\alpha+\mathbf{D}_t(1-\psi)^{-1}\mu^\gamma=(\mathbf{I}_N-\overline\psi\mathbf{G}_t)^{-1}(\mathbf{I}_N+\overline\rho\mathbf{G}_t)\overline\mu^\alpha+\mathbf{D}_t(1-\overline\psi)^{-1}\overline\mu^\gamma
\end{align}
Pre-multiplying both sides by $(\mathbf{I}_N-\psi\mathbf{G}_t)(\mathbf{I}_N-\overline\psi\mathbf{G}_t)$ and rearranging yields
\begin{align}
&\mathbf{I}_N(\mu^\alpha-\overline\mu^\alpha)+\mathbf{G}_t\left((\rho-\overline\psi)\mu^\alpha-(\overline\rho-\psi)\overline\mu^\alpha\right)+\mathbf{G}_t^2\left(-\overline\psi\rho\mu^\alpha+\psi\overline\rho\overline\mu^\alpha\right)\nonumber\\
&+\mathbf{D}_t\left((1-\overline\psi)\mu^\gamma-(1-\psi)\overline\mu^\gamma  \right)=\mathbf{0}
\end{align}
Stacking for $t\in[T]$ yields
\begin{align}
\mathbf{J}(\mu^\alpha-\overline\mu^\alpha)+\mathbf{G}\left((\rho-\overline\psi)\mu^\alpha-(\overline\rho-\psi)\overline\mu^\alpha\right)+\mathbf{H}\left(-\overline\psi\rho\mu^\alpha+\psi\overline\rho\overline\mu^\alpha\right)+\mathbf{D}\left((1-\overline\psi)\mu^\gamma-(1-\psi)\overline\mu^\gamma  \right)=\mathbf{0}
\end{align}
and applying $\mathbf{W}$ on the left yields
\begin{align}
[\mathbf{WJ},\mathbf{WG},\mathbf{WH}]\begin{pmatrix}\mu^\alpha-\overline\mu^\alpha\\(\rho-\overline\psi)\mu^\alpha-(\overline\rho-\psi)\overline\mu^\alpha \\ -\overline\psi\rho\mu^\alpha+\psi\overline\rho\overline\mu^\alpha  \end{pmatrix}=\mathbf{0}
\end{align}
Let $\mathbf{v}=(\mathbf{v}_1',\mathbf{v}_2',\mathbf{v}_3')$ be a vector in the null-space of $[\mathbf{WJ},\mathbf{WG},\mathbf{WH}]$. Then for some $\mathbf{v}$,
\begin{align}
&\mathbf{v}_1=\mu^\alpha-\overline\mu^\alpha,\quad\mathbf{v}_2=(\rho-\overline\psi)\mu^\alpha-(\overline\rho-\psi)\overline\mu^\alpha,\quad\mathbf{v}_3=-\overline\psi\rho\mu^\alpha+\psi\overline\rho\overline\mu^\alpha
\end{align}
Eliminating $\overline\mu^\alpha$ yields
\begin{align}
&(\overline\rho-\psi)\mathbf{v}_1-\mathbf{v}_2=(\overline\rho-\psi+\overline\psi-\rho)\mu^\alpha,\quad\psi\overline\rho\mathbf{v}_1+\mathbf{v}_3=(\psi\overline\rho-\overline\psi\rho)\mu^\alpha
\end{align}
If there does not exist $\mathbf{v}$ in the null-space of $[\mathbf{WJ},\mathbf{WG},\mathbf{WH}]$ and scalars $\delta_1,\delta_2,\delta_3,\delta_4$ satisfying $\delta_1\mathbf{v}_1-\mathbf{v}_2=(\delta_1-\delta_2)\mu^\alpha$, $\delta_3\mathbf{v}_1+\mathbf{v}_3=(\delta_3-\delta_4)\mu^\alpha$ and either $\delta_1\neq \delta_2$ or $\delta_3\neq\delta_4$ then we have
\begin{align}
\overline\rho-\psi+\overline\psi-\rho=0,\quad \psi\overline\rho-\overline\psi\rho=0
\end{align}
Solving the first equation for $\overline\psi$ and injecting into the second yields $(\overline\rho-\rho)(\psi+\rho)=0$, hence $\overline\rho=\rho$ provided that $\psi+\rho\neq 0$. Injecting $\overline\rho=\rho$ into the first equation yields $\overline\psi=\psi$. 

\noindent\textbf{Proof of Proposition \ref{the2y}}

\noindent Denote $\theta=(\psi,\rho,\sigma^2_\alpha)'$. We have the reduced form
\begin{align}
\mathbf{y}=(\mathbf{I}_{NT}-\psi\mathbf{F})^{-1}(\mathbf{J}+\rho\mathbf{G})\alpha+\mathbf{D}(1-\psi)^{-1}\beta+(\mathbf{I}_{NT}-\psi\mathbf{F})^{-1}\epsilon
\end{align}
Since $\mathbf{G}=\overline{\mathbf{G}}$, we have $\mathbf{G}_t=\mathbf{G}_t^2$ for $t\in[T]$ and $\mathbf{F}=\mathbf{F}^2$, hence $(\mathbf{I}_{NT}-\psi\mathbf{F})^{-1}=\mathbf{I}_{NT}+\frac{\psi}{1-\psi}\mathbf{F}$ and
\begin{align}
\mathbf{W}\mathbf{y}=\mathbf{W}\left(\mathbf{J}+\frac{\psi+\rho}{1-\psi}\mathbf{G}\right)\alpha+\mathbf{W}\left(\mathbf{I}_{NT}+\frac{\psi}{1-\psi}\mathbf{F}\right)\epsilon
\end{align}
with variance
\begin{align}
\mathbb{V}[\mathbf{Wy}]&=\sigma^2_\alpha\mathbf{W}\left(\mathbf{J}+\frac{\psi+\rho}{1-\psi}\mathbf{G}\right)\left(\mathbf{J}+\frac{\psi+\rho}{1-\psi}\mathbf{G}\right)\mathbf{W}\\
&+\mathbf{W}\left(\mathbf{I}_{NT}+\frac{\psi}{1-\psi}\mathbf{F}\right)\Sigma\left(\mathbf{I}_{NT}+\frac{\psi}{1-\psi}\mathbf{F}\right)\mathbf{W}\\
&=\sigma^2_\alpha\mathbf{WJJ'W}+\frac{\sigma^2_\alpha(\psi+\rho)}{1-\psi}\mathbf{W}(\mathbf{JG}'+\mathbf{GJ}')\mathbf{W}+\frac{\sigma^2_\alpha(\psi+\rho)^2}{(1-\psi)^2}\mathbf{W}\mathbf{GG}'\mathbf{W}\\
&+\mathbf{W}\Sigma\mathbf{W}+\frac{\psi(2-\psi)}{(1-\psi)^2}\mathbf{W}\mathbf{F}\Sigma\mathbf{F}\mathbf{W}
\end{align}
where $\Sigma=\mathbb{V}[\epsilon]$. Consider group $m\in[M]$ and define $\mathcal{E}_m\subseteq[NT]$ to be the indices of the rows of $\mathbf{y}$ which correspond to group $m$. Then we can define $\mathbf{E}_m$ as the matrix constructed from rows $\mathcal{E}_m$ of $\mathbf{I}_{NT}$. Pre-multiplying any conformable matrix by $\mathbf{E}_m$ extracts rows $\mathcal{E}_m$, and post-multiplying by $\mathbf{E}_m'$ extracts columns $\mathcal{E}_m$. Consider first the within-group variance for group $m$, given by
\begin{align}
&\sigma^2_\alpha\mathbf{E}_m\mathbf{WJJ'W}\mathbf{E}_m'+\frac{\sigma^2_\alpha(\psi+\rho)}{1-\psi}\mathbf{E}_m\mathbf{W}(\mathbf{JG}'+\mathbf{GJ}')\mathbf{W}\mathbf{E}_m'\\
&+\frac{\sigma^2_\alpha(\psi+\rho)^2}{(1-\psi)^2}\mathbf{E}_m\mathbf{W}\mathbf{GG}'\mathbf{W}\mathbf{E}_m'+\sigma^2(m)\mathbf{E}_m\mathbf{W}\mathbf{E}_m'+\frac{\sigma^2(m)\psi(2-\psi)}{(1-\psi)^2}\mathbf{E}_m\mathbf{W}\mathbf{F}\mathbf{W}\mathbf{E}_m'
\end{align}
and the between group variance given by
\begin{align}
&\sigma^2_\alpha\mathbf{E}_m\mathbf{WJJ'W}\mathbf{E}_{-m}'+\frac{\sigma^2_\alpha(\psi+\rho)}{1-\psi}\mathbf{E}_m\mathbf{W}(\mathbf{JG}'+\mathbf{GJ}')\mathbf{W}\mathbf{E}_{-m}'+\frac{\sigma^2_\alpha(\psi+\rho)^2}{(1-\psi)^2}\mathbf{E}_m\mathbf{W}\mathbf{GG}'\mathbf{W}\mathbf{E}_{-m}'
\end{align}
where $\mathbf{E}_{-m}$ extracts the rows $\mathcal{E}_m^c$, which denotes the complement of $\mathcal{E}_{m}$. Notice first that the between-group variance alone is insufficient to identify $\psi$ and $\rho$. Only $(\psi+\rho)(1-\psi)^{-1}$ is identifiable based on $\mathbb{COV}[\epsilon_{it},\epsilon_{jt}]=0$ for all $(i,j,t)\in[N]^2\times T:g(i,t)\neq g(j,t)$ alone. However, under the additional restrictions on the within-group variance in \eqref{var}, we obtain two additional terms which allow additionally for identification of $\psi(2-\psi)(1-\psi)^{-2}$. Now suppose that there is $\overline\theta$ satisfying the conditional variance restrictions in \eqref{var}. Then, if $\mathbf{W},\mathbf{WJJ'W},\mathbf{W}(\mathbf{JG}'+\mathbf{GJ}')\mathbf{W},\mathbf{WGG'W}$ and $\mathbf{WFW}$ are linearly independent then there exists $m$ such that $\overline\sigma^2(m)=\sigma^2(m)$, $\overline\sigma^2_\alpha=\sigma^2_\alpha$ and
\begin{align}
&\frac{\psi+\rho}{1-\psi}=\frac{\overline\psi+\overline\rho}{1-\overline\psi},\quad\frac{\psi(2-\psi)}{(1-\psi)^2}=\frac{\overline\psi(2-\overline\psi)}{(1-\overline\psi)^2}
\end{align}
These two equations have two solutions given by $\overline\psi=\psi,\overline\rho=\rho$ and $\overline\psi=2-\psi,\overline\rho=(2\psi+3\rho-\psi\rho-2)(1+\psi)^{-1}$. Since $|\psi|<1$ we have $|2-\psi|>1$, hence the second solution is infeasible. Hence we obtain $\overline\theta=\theta$.

\noindent\textbf{Proof of Proposition \ref{the3y}} Denote $\theta=(\psi,\rho,\sigma^2_\alpha)'$.  Under the variance restriction \eqref{var} we have
\begin{align}
\mathbb{V}[\mathbf{y}]&=\sigma^2_\alpha(\mathbf{I}_{NT}-\psi\mathbf{F})^{-1}(\mathbf{J}+\rho\mathbf{G})(\mathbf{J}+\rho\mathbf{G})'(\mathbf{I}_{NT}-\psi\mathbf{F}')^{-1}+\frac{1}{(1-\psi)^2}\mathbf{D}\Sigma_\gamma\mathbf{D}'\\
&+(\mathbf{I}_{NT}-\psi\mathbf{F})^{-1}\Sigma(\mathbf{I}_{NT}-\psi\mathbf{F}')^{-1}+\frac{1}{1-\psi}(\mathbf{I}_{NT}-\psi\mathbf{F})^{-1}(\mathbf{J}+\rho\mathbf{G})\Sigma_{\alpha\gamma}\mathbf{D}'\\
&+\frac{1}{1-\psi}\mathbf{D}\Sigma_{\alpha\gamma}'(\mathbf{J}+\rho\mathbf{G})'(\mathbf{I}_{NT}-\psi\mathbf{F}')^{-1}+\frac{1}{1-\psi}(\mathbf{I}_{NT}-\psi\mathbf{F})^{-1}\Sigma_{\epsilon\gamma}\mathbf{D}'\\
&+\frac{1}{1-\psi}\mathbf{D}\Sigma_{\epsilon\gamma}'(\mathbf{I}_{NT}-\psi\mathbf{F}')^{-1}
\end{align}
where $\Sigma$ encodes the variance of $\epsilon$, $\Sigma_{\alpha\gamma}$ encodes the covariance terms for $\alpha,\gamma$ and similarly for $\Sigma_{\epsilon\gamma}$. Suppose that there is $\overline\theta$ satisfying the variance restriction in \eqref{vargen}. Then we have
\begin{align}
&\sigma^2_\alpha(\mathbf{I}_{NT}-\psi\mathbf{F})^{-1}(\mathbf{J}+\rho\mathbf{G})(\mathbf{J}+\rho\mathbf{G})'(\mathbf{I}_{NT}-\psi\mathbf{F}')^{-1}+\frac{1}{(1-\psi)^2}\mathbf{D}\Sigma_\gamma\mathbf{D}'\\
&+(\mathbf{I}_{NT}-\psi\mathbf{F})^{-1}\Sigma(\mathbf{I}_{NT}-\psi\mathbf{F}')^{-1}+\frac{1}{1-\psi}(\mathbf{I}_{NT}-\psi\mathbf{F})^{-1}(\mathbf{J}+\rho\mathbf{G})\Sigma_{\alpha\gamma}\mathbf{D}'\\
&+\frac{1}{1-\psi}\mathbf{D}\Sigma_{\alpha\gamma}'(\mathbf{J}+\rho\mathbf{G})'(\mathbf{I}_{NT}-\psi\mathbf{F}')^{-1}+\frac{1}{1-\psi}(\mathbf{I}_{NT}-\psi\mathbf{F})^{-1}\Sigma_{\epsilon\gamma}\mathbf{D}'\\
&+\frac{1}{1-\psi}\mathbf{D}\Sigma_{\epsilon\gamma}'(\mathbf{I}_{NT}-\psi\mathbf{F}')^{-1}=\\
&\overline{\sigma}^2_\alpha(\mathbf{I}_{NT}-\overline\psi\mathbf{F})^{-1}(\mathbf{J}+\overline\rho\mathbf{G})(\mathbf{J}+\overline\rho\mathbf{G})'(\mathbf{I}_{NT}-\overline\psi\mathbf{F}')^{-1}+\frac{1}{(1-\overline\psi)^2}\mathbf{D}\overline\Sigma_\gamma\mathbf{D}'\\
&+\epsilon(\mathbf{I}_{NT}-\overline\psi\mathbf{F})^{-1}\overline\Sigma(\mathbf{I}_{NT}-\overline\psi\mathbf{F}')^{-1}+\frac{1}{1-\overline\psi}(\mathbf{I}_{NT}-\overline\psi\mathbf{F})^{-1}(\mathbf{J}+\overline\rho\mathbf{G})\overline\Sigma_{\alpha\gamma}\mathbf{D}'\\
&+\frac{1}{1-\overline\psi}\mathbf{D}\overline\Sigma_{\alpha\gamma}'(\mathbf{J}+\overline\rho\mathbf{G})'(\mathbf{I}_{NT}-\overline\psi\mathbf{F}')^{-1}+\frac{1}{1-\overline\psi}(\mathbf{I}_{NT}-\overline\psi\mathbf{F})^{-1}\overline\Sigma_{\epsilon\gamma}\mathbf{D}'\\
&+\frac{1}{1-\overline\psi}\mathbf{D}\overline\Sigma_{\epsilon\gamma}'(\mathbf{I}_{NT}-\overline\psi\mathbf{F}')^{-1}
\end{align}
Pre-multiplying both sides by $\mathbf{W}(\mathbf{I}_{NT}-\psi\mathbf{F})(\mathbf{I}_{NT}-\overline\psi\mathbf{F})$ and post-multiplying by $(\mathbf{I}_{NT}-\psi\mathbf{F}')(\mathbf{I}_{NT}-\overline\psi\mathbf{F}')\mathbf{W}$ yields
\begin{align}
&\sigma^2_\alpha\mathbf{W}(\mathbf{I}_{NT}-\overline\psi\mathbf{F})(\mathbf{J}+\rho\mathbf{G})(\mathbf{J}+\rho\mathbf{G})'(\mathbf{I}_{NT}-\overline\psi\mathbf{F}')\mathbf{W}+\mathbf{W}(\mathbf{I}_{NT}-\overline\psi\mathbf{F})\Sigma(\mathbf{I}_{NT}-\overline\psi\mathbf{F}')\mathbf{W}=\nonumber\\
&\overline{\sigma}^2_\alpha\mathbf{W}(\mathbf{I}_{NT}-\psi\mathbf{F})(\mathbf{J}+\overline\rho\mathbf{G})(\mathbf{J}+\overline\rho\mathbf{G})'(\mathbf{I}_{NT}-\psi\mathbf{F}')\mathbf{W}+\mathbf{W}(\mathbf{I}_{NT}-\psi\mathbf{F})\overline\Sigma(\mathbf{I}_{NT}-\psi\mathbf{F}')\mathbf{W}
\end{align}
Rearranging yields
\begin{align}
&(\sigma^2_\alpha-\overline\sigma^2_\alpha)\mathbf{WJJ'W}+(\sigma^2_\alpha(\rho-\overline\psi)-\overline\sigma^2_\alpha(\overline\rho-\psi))\mathbf{W}(\mathbf{JG}'+\mathbf{GJ}')\mathbf{W}\\
&+(\overline\sigma^2_\alpha\psi\overline\rho-\sigma^2_\alpha\overline\psi\rho)\mathbf{W}(\mathbf{JH}'+\mathbf{HJ}')\mathbf{W}+(\overline\sigma^2_\alpha(\overline\rho-\psi)\psi\overline\rho-\sigma^2_\alpha(\rho-\overline\psi)\overline\psi\rho)\mathbf{W}(\mathbf{GH}'+\mathbf{HG}')\mathbf{W}\\
&+(\sigma^2_\alpha(\rho-\overline\psi)^2-\overline\sigma^2_\alpha(\overline\rho-\psi)^2)\mathbf{WGG'W}+(\sigma^2_\alpha\overline\psi^2\rho^2-\overline\sigma^2_\alpha\psi^2\overline\rho^2)\mathbf{WHH'W}+\mathbf{W}(\Sigma-\overline\Sigma)\mathbf{W}\\
&+\mathbf{W}(\mathbf{F}(\psi\overline\Sigma-\overline\psi\Sigma)+(\psi\overline\Sigma-\overline\psi\Sigma)\mathbf{F}')\mathbf{W}+\mathbf{WF}(\overline\psi^2\Sigma-\psi^2\overline\Sigma)\mathbf{F'W}=\mathbf{0}
\end{align}
Since the within-group variance in $\epsilon$ is not restricted by \eqref{vargen}, identification hinges on between-group variance in the outcomes. Consider the covariance terms for the outcomes of group $m$ with those in all other groups, for which we have the equations
\begin{align}
&(\sigma^2_\alpha-\overline\sigma^2_\alpha)  \mathbf{E}_m\mathbf{WJJ'W}\mathbf{E}_{-m}'+(\sigma^2_\alpha(\rho-\overline\psi)-\overline\sigma^2_\alpha(\overline\rho-\psi))  \mathbf{E}_m\mathbf{W}(\mathbf{JG}'+\mathbf{GJ}')\mathbf{W}\mathbf{E}_{-m}'\nonumber\\
&+(\overline\sigma^2_\alpha\psi\overline\rho-\sigma^2_\alpha\overline\psi\rho)\mathbf{E}_m\mathbf{W}(\mathbf{JH}'+\mathbf{HJ}')\mathbf{W}\mathbf{E}_{-m}'\\
&+(\overline\sigma^2_\alpha(\overline\rho-\psi)\psi\overline\rho-\sigma^2_\alpha(\rho-\overline\psi)\overline\psi\rho)\mathbf{E}_m\mathbf{W}(\mathbf{GH}'+\mathbf{HG}')\mathbf{W}\mathbf{E}_{-m}'\\
&+(\sigma^2_\alpha(\rho-\overline\psi)^2-\overline\sigma^2_\alpha(\overline\rho-\psi)^2)\mathbf{E}_m\mathbf{WGG'W}\mathbf{E}_{-m}'\\
&+(\sigma^2_\alpha\overline\psi^2\rho^2-\overline\sigma^2_\alpha\psi^2\overline\rho^2)\mathbf{E}_m\mathbf{WHH'W}\mathbf{E}_{-m}'=\mathbf{0}
\end{align}
where $\mathbf{E}_{m}$ and $\mathbf{E}_{-m}'$ are defined in the proof of Proposition \ref{the2y}. If there exists $m\in[M]$ such that $\mathbf{E}_{m}\mathbf{WJJ'W}\mathbf{E}_{-m}',  \mathbf{E}_m\mathbf{W}(\mathbf{JG}'+\mathbf{GJ}')\mathbf{W}\mathbf{E}_{-m}'$ and $\mathbf{E}_m\mathbf{W}(\mathbf{JH}'+\mathbf{HJ}')\mathbf{W}\mathbf{E}_{-m}'$ are maximially linearly independent from the other matrices then we have $\sigma^2_\alpha=\overline\sigma^2_\alpha$ and
\begin{align}
\overline\rho-\psi+\overline\psi-\rho=0\label{ey11},\quad \psi\overline\rho-\overline\psi\rho=0
\end{align}
Solving the first equation for $\overline\psi$ and injecting into the second yields $(\overline\rho-\rho)(\psi+\rho)=0$, hence $\overline\rho=\rho$ provided that $\psi+\rho\neq 0$. Injecting into the first equation yields $\overline\psi=\psi$.

\end{document}